\documentclass[12pt]{spieman}
\usepackage{graphicx}
\usepackage{tocloft}
\usepackage{mathabx}
\usepackage{eucal}
\usepackage[caption=false]{subfig}
\usepackage{eucal}
\usepackage{setspace}
\usepackage[left]{lineno}
\newcommand\arcdeg{\mbox{$^\circ$}}
\newcommand\arcmin{\mbox{$^\prime$}}
\newcommand\arcsec{\mbox{$^{\prime\prime}$}}

\title{AIROPA III: Testing Simulated and On-Sky Data}

\author[a,b,*]{Paolo Turri}
\author[c]{Jessica R. Lu}
\author[d]{Gunther Witzel}
\author[e]{Anna Ciurlo}
\author[e]{Tuan Do}
\author[e]{Andrea M. Ghez}
\author[e]{Michael P. Fitzgerald}
\author[f]{Matthew C. Britton}
\author[g]{Sam Ragland}
\author[c]{Sean K. Terry}

\affil[a]{University of British Columbia, Department of Physics \& Astronomy, 6224 Agricultural Rd, Vancouver, BC, Canada, V6T 1Z1}
\affil[b]{National Research Council, Herzberg Astronomy and Astrophysics, 5071 W. Saanich Rd, Victoria, BC, Canada, V9E 2E7}
\affil[c]{University of California - Berkeley, Astronomy Department, 501 Campbell Hall, Berkeley, CA, USA, 94720}
\affil[d]{Max Planck Institute for Radio Astronomy, Auf dem H\"{u}gel 69, Bonn, Germany, D-53121}
\affil[e]{University of California - Los Angeles, Division of Astronomy \& Astrophysics, Los Angeles, CA, USA, 90095}
\affil[f]{The Aerospace Corporation, 2310 E. El Segundo Blvd, El Segundo, CA, USA, 90245}
\affil[g]{W. M. Keck Observatory, 65-1120 Mamalahoa Hwy, Kamuela, HI, USA, 96743}

\cftpagenumbersoff{figure}
\cftpagenumbersoff{table} 
\begin{document}
\maketitle

\begin{abstract}

Adaptive optics images from the W. M. Keck Observatory have delivered numerous influential scientific results, including detection of multi-system asteroids, the supermassive black hole at the center of the Milky Way, and directly imaged exoplanets.
Specifically, the precise and accurate astrometry these images yield was used to measure the mass of the supermassive black hole using orbits of the surrounding star cluster.
Despite these successes, one of the major obstacles to improved astrometric measurements is the spatial and temporal variability of the point-spread function delivered by the instruments.
AIROPA is a software package for the astrometric and photometric analysis of adaptive optics images using point-spread function fitting together with the technique of point-spread function reconstruction.
In adaptive optics point-spread function reconstruction, the knowledge of the instrument performance and of the atmospheric turbulence is used to predict the long-exposure point-spread function of an observation.
In this paper we present the results of our tests using AIROPA on both simulated and on-sky images of the Galactic Center.
We find that our method is very reliable in accounting for the static aberrations internal to the instrument, but it does not improve significantly the accuracy on sky, possibly due to uncalibrated telescope aberrations.
\end{abstract}

\keywords{adaptive optics, PSF reconstruction, Galactic Center, astrometry}

{\noindent\footnotesize\textbf{*}Paolo Turri, \linkable{turri@astro.ubc.ca}}

\section{Introduction}

Adaptive optics (AO) is a technology used in ground-based optical and near-infrared (NIR) astronomy to compensate for the blurring effects of the Earth's atmosphere (see Ref.~\citenum{bib:davies12} for a comprehensive review on the subject).
Atmospheric turbulence deteriorates the flat incoming wavefront of light and the corresponding diffraction-limited point-spread function (PSF) of a telescope into a broad, seeing-limited PSF, with a width set by the amount of turbulence.
AO corrects the aberrations using wavefront sensing cameras and deformable mirrors operating at milli-second timescales.
By restoring the diffraction-limited full width at half maximum (FWHM) of the PSF, AO achieves observations with a higher spatial resolution and signal-to-noise ratio than seeing-limited observations.

One of the most successful applications of AO in astronomy has been the study of the Galactic Center (GC) at the W. M. Keck Observatory.
The compact radio source Sgr A\textsuperscript{*} at the center of our Galaxy \cite{bib:balick74} has a NIR counterpart surrounded by a cluster of high proper motion, orbiting stars \cite{bib:ghez98,bib:ghez00,bib:genzel10}.
AO has been used to resolve this dense environment, and make precise astrometric, photometric, and spectroscopic measurements of individual stars for the past 20 years.
These measurements have been used to prove that a supermassive black hole (SMBH) resides at the center with a mass of \mbox{4.02 $\times$ 10\textsuperscript{6} $\,M_{\Sun}$} at a distance of 8.0 kpc \cite{bib:ghez05,bib:ghez08,bib:boehle16}, surrounded by a young nuclear star cluster whose formation is still not well understood, an old nuclear star cluster with an unusual metallicity distribution, and short-period orbiting stars \cite{bib:lu09,bib:do13} that have been used to test general relativity and other theories of gravity \cite{bib:hees17}.

One of the primary limitations to studies with AO of crowded fields are the systematic errors caused by the imperfect knowledge of the PSF.
The accuracy achievable by AO instrumentation in astrometry, photometry and the measurement of spectra is degraded if the PSF model used for fitting is not representative of the data \cite{bib:schodel10,bib:trippe10,bib:ascenso15,bib:fritz15,bib:turri17}.
One of the most difficult aspects of the PSF to model is the spatial variability over the field of view.
The main sources of field-dependent PSF variation are uncorrected atmospheric turbulence \cite{bib:fusco00} and non-common path aberrations \cite{bib:lamb14}.
In single-conjugate AO systems, like the one at Keck, the correction is optimized in the direction of the guide star and deteriorates rapidly outside a radius of several arcseconds, an effect known as angular anisoplanatism \cite{bib:fried82}.
This is caused by the loss of correlation between the wavefront measured in the direction of the guide stars and the wavefront corrected in a different direction.
The effect is the elongation of the PSF in the direction of the guide star.
The magnitude of the angular anisoplanatism depends on the distance from the guide star, on the vertical distribution of the atmospheric turbulence, and on the elevation angle of the telescope.

One solution to model a field-dependent PSF is to use many bright and isolated stars to measure it at different field positions \cite{bib:stetson87,bib:schreiber12}.
The more the PSF varies, the larger the required number of these ``PSF stars''.
In the GC field of view of NIRC2 and OSIRIS, two of the instruments served by the Keck AO system, there are only a dozen of suitable PSF stars, not enough to characterize the observed spatial variation.
In addition, the extreme degree of crowding in the region does not allow a clear measurement of the profile of most of them.
With just a limited number of sources available, so far only a constant PSF has been used for PSF fitting of the Keck observations.

An alternative to the direct measurement of the PSF is to predict it by modelling the optical system and the atmospheric turbulence using the technique of PSF-reconstruction \cite{bib:veran97,bib:wagner19}.
While this approach has been proved both in theory and on sky \cite{bib:martin16,bib:ragland18a,bib:gilles18,bib:massari20}, it has not been used consistently for scientific observations.
Furthermore, the off-axis variations of the PSF are not predicted and must be modeled.

AIROPA (Anisoplanatic and Instrumental Reconstruction of Off-axis PSFs for AO) is a PSF-fitting software with field-dependent PSF-reconstruction, developed with the goal of improving the accuracy of the GC astrometry and photometry with NIRC2, the Keck AO imager.
AIROPA models two components of the PSF variability--the instrumental aberrations and the atmospheric aberrations--and can be used to analyze NIRC2 and OSIRIS images, which are the two instruments fed by the Keck AO systems.

AIROPA is described in a series of papers: an overview \cite{bib:witzel16}, details of the instrumental modeling \cite{bib:ciurlo21}, and on-sky testing in a wide range of atmospheric conditions \cite{bib:terry21}.
In this paper we discuss the testing of AIROPA with simulated and on-sky NIRC2 images.
In Section~\ref{sec:airopa}, we outline the operating principles of AIROPA on NIRC2 images.
The instrumental and astronomical data used for this paper are introduced in Section~\ref{sec:data}.
In Section~\ref{sec:sim} we present the astrometric and photometric analysis with AIROPA of several simulated images, while in Section~\ref{sec:sky} we use our software on real on-sky images of the GC taken with NIRC2.
Section~\ref{sec:summary} discusses the results of our tests of AIROPA's performance.

\section{AIROPA}\label{sec:airopa}

AIROPA reconstructs the PSF for an image through a combination of (1) empirical extraction from stars in the image and (2) model prediction from atmospheric turbulence profiles and instrumental aberration maps.
The AIROPA modules for star detection, empirical PSF extraction, and PSF fitting are built upon StarFinder \cite{bib:diolaiti00}, an IDL program that extracts stellar astrometry and photometry from focal plane images using a single PSF over the whole image.
AIROPA improves on StarFinder by building and fitting a grid of reconstructed, spatially variable PSFs defined across the FOV.
It also uses an improved algorithm to smooth the PSF halo and to make it converge to zero at the edges, by clipping the values that are below the noise level and that are not contiguous to the rest of the PSF model.

\begin{figure}
\centering
\includegraphics[width=0.7\textwidth]{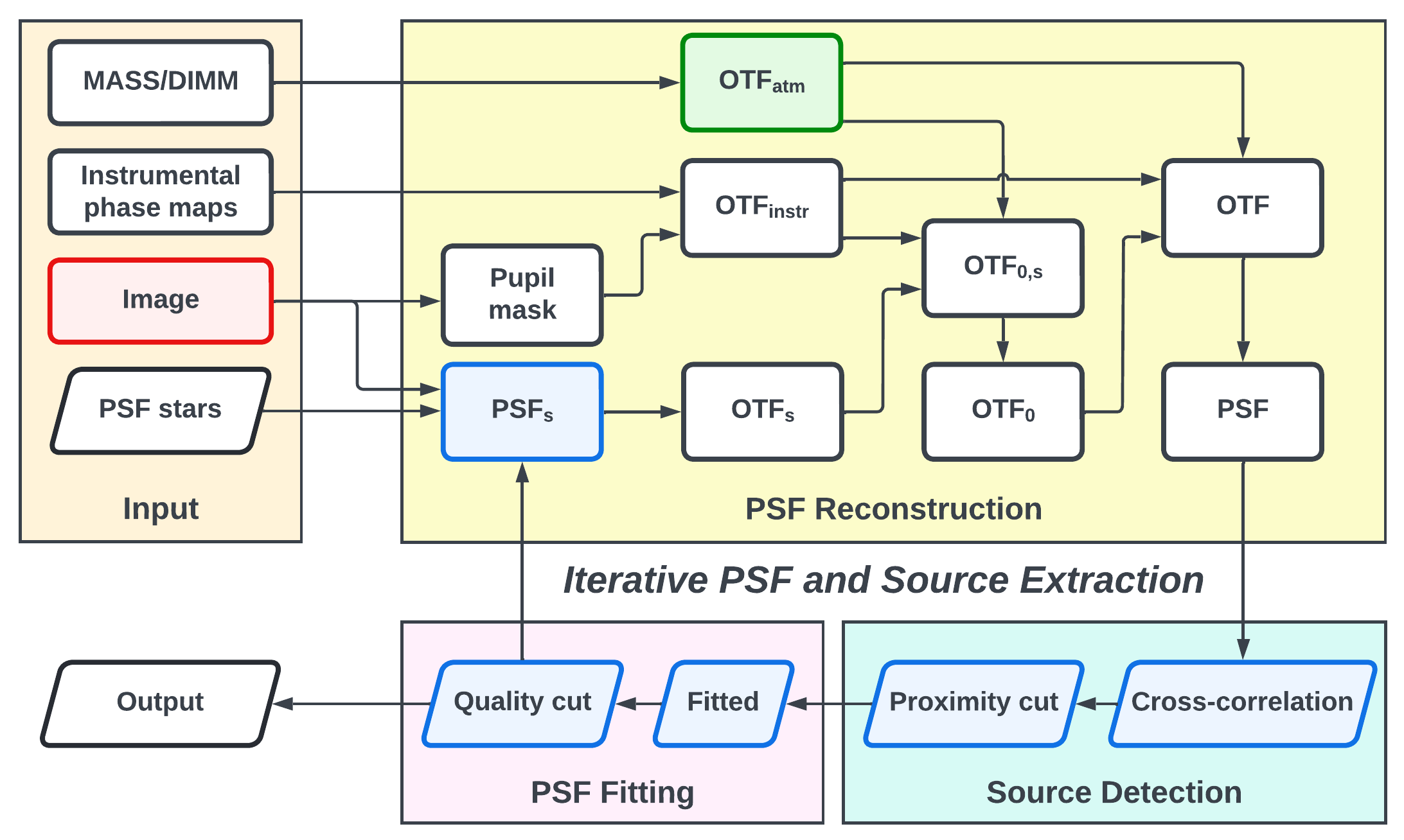}
\caption{Diagram of AIROPA in \texttt{variable-PSF} mode.
The elements represented by a parallelogram are star catalogs.
Blue items are produced by our modified version of StarFinder, green by ARROYO, and red by the NIRC2 image reduction pipeline.\label{fig:airopa}}
\end{figure}

AIROPA offers two alternative approaches to PSF modelling.
The classical \texttt{single-PSF} mode uses StarFinder to fit every star in an image using the same PSF model, empirically extracted from the median of the normalized profiles of a user-selected set of PSF reference stars.
The \texttt{variable-PSF} algorithm (Figure \ref{fig:airopa}) uses instead a grid of different PSFs across the field of view, built from a combination of the empirically extracted PSF and a field-dependent model of anisoplanatic and instrumental aberrations\cite{bib:witzel16}.

The PSF of an exposure taken at the time $t$, and at the position $\mathbf{r}$ in the NIRC2 FOV, can be described as the convolution of the on-axis ($\mathbf{r}=0$) PSF with an instrumental and atmospheric component that characterize its spatial variability:
\begin{equation}
PSF(\mathbf{r},t)=PSF_{0}(t)*PSF_{inst}(\mathbf{r},t)*PSF_{atm}(\mathbf{r},t).
\end{equation}
For faster computation and easier manipulation of the terms, we use the convolution theorem with the optical transfer function (OTF), defined as the Fourier transform of the PSF:
\begin{equation}
OT\!F(\mathbf{r},t)=OT\!F_{0}(t)\cdot OT\!F_{inst}(\mathbf{r},t)\cdot OT\!F_{atm}(\mathbf{r},t).\label{eq:otf}
\end{equation}
The $OT\!F_{inst}$ and $OT\!F_{atm}$ are the ratios of the OTF at the position $\mathbf{r}$ respect to the one on-axis $OT\!F_{0}$, caused by instrumental aberrations and AO angular anisoplanatism, respectively.
The OTF is composed of a real part (called the modulation transfer function, or MTF) and an imaginary part (called the phase transfer function, or PhTF).

The NIRC2 contribution to the OTF can be considered constant with time during one year of observations of the GC (see Section~\ref{sec:calib}), except for the position angle of the telescope pupil as seen by the instrument, which changes between exposures as the telescope tracks the target during the night.
$OT\!F_{inst}$ is measured typically once a year at several positions in the NIRC2 FOV (see Section~\ref{sec:inst}).
The OTF ratio is then masked using the telescope pupil rotated by the position angle it had at the time of the exposure \cite{bib:ciurlo21}, recorded in the NIRC2 FITS image header.

The atmospheric factor $OT\!F_{atm}$ is calculated on a grid within the FOV using the ARROYO code \cite{bib:britton06}, which models the angular anisoplanatism produced by the Keck AO system.
This program requires knowledge of the position of the natural guide star (NGS) and laser guide star (LGS) relative to the frame, as well as the turbulence profile measured at the time of the exposure (Section ~\ref{sec:atm}).

Last, the spatially constant $OT\!F_{0}$ of the exposure is extracted empirically from the PSF stars.
Each star's individual, off-axis $PSF_{s}$ is Fourier-transformed into $OT\!F_{s}=\mathcal{F}\{PSF_{s}\}$. Following Eq.~\ref{eq:otf}, the on-axis $OT\!F_{0,s}$ of a star can be computed by removing the instrumental and atmospheric components:
\begin{equation}
OT\!F_{0,s}(t)=\frac{OT\!F_{s}(\mathbf{r},t)}{OT\!F_{inst}(\mathbf{r},t)\cdot OT\!F_{atm}(\mathbf{r},t)}.
\end{equation}
The $OT\!F_{0}$ of an image is the average of the $OT\!F_{0,s}$ from the set of selected PSF reference stars.

Once all three OTF components are in place, the $OT\!F(\mathbf{r},t)$ for a position in an exposure is obtained using Eq.~\ref{eq:otf}, and the $PSF(\mathbf{r},t)$ is calculated as its inverse Fourier transform.
The final PSF is cut to a diameter of 150 px (1.5\arcsec), since the reconstruction beyond this distance is imprecise, given the typical exposure time, brightness and number of PSF stars of our observations.

StarFinder cross-correlates the reconstructed PSF with the image to identify the stars in the field.
Some of them are spurious detections of speckles, and are removed from the list, based on the magnitude difference and proximity to other stars.
The catalog is then used to fit the reconstructed PSF by minimizing
the least squares error between the data and the model.
Objects with poor fittings, such as galaxies or cosmic rays, are dismissed.
The catalog so obtained can be used to subtract from the image the sources close to the PSF stars, yielding a more accurate PSF and detections.
For our analysis, we repeat this iteration three times before reaching the final catalog of sources.

\section{Data to Validate AIROPA}\label{sec:data}

\subsection{Instrument Description}

The Keck LGS AO System consists of a pair of similar single-conjugate AO instruments mounted on a Nasmyth platform of both Keck I and II telescopes \cite{bib:wizinowich06}.
It employs both an artificial LGS and a natural star for measuring the tip-tilt modes only (T/T star).
Each system delivers a correction in the NIR using a single deformable mirror and a tip-tilt mirror to correct the aberrations.
The two systems serve different scientific instruments.
For this paper we concentrate on the system on Keck II that serves the NIRC2 imager.
Its 1024$\times$1024 pixels cover a field of view of 10.2\arcsec with a scale of 9.942 mas px$^\mathrm{-1}$ \cite{bib:yelda10}.
The detector is an Aladdin III InSb with a gain of 4.0 $\mathrm{e^-}$ ADU$^\mathrm{-1}$ and a dark current of 0.1 $\mathrm{e^-}$ px$^\mathrm{-1}$ s$^\mathrm{-1}$.
NIRC2's read noise is 60.0 $\mathrm{e^-}$ when using a Fowler sampling \cite{bib:fowler90} of 8 (the setting typically used with NIRC2).

\subsection{Phase Maps to Calibrate Instrumental Aberrations}\label{sec:inst}

The instrumental aberrations of the AO system and imager are characterized by phase maps measured using out-of-focus images of a fiber source at 81 positions (on a 9$\times$9 grid) in the NIRC2 field of view \cite{bib:ciurlo18,bib:ciurlo21,bib:sitarski14}.
The phase maps were recovered from the fiber images by using the Gerchberg-Saxton algorithm \cite{bib:gerchberg72}, the same method used for the image sharpening of the instrument.

\begin{figure}
\centering
\includegraphics[width=0.9\textwidth]{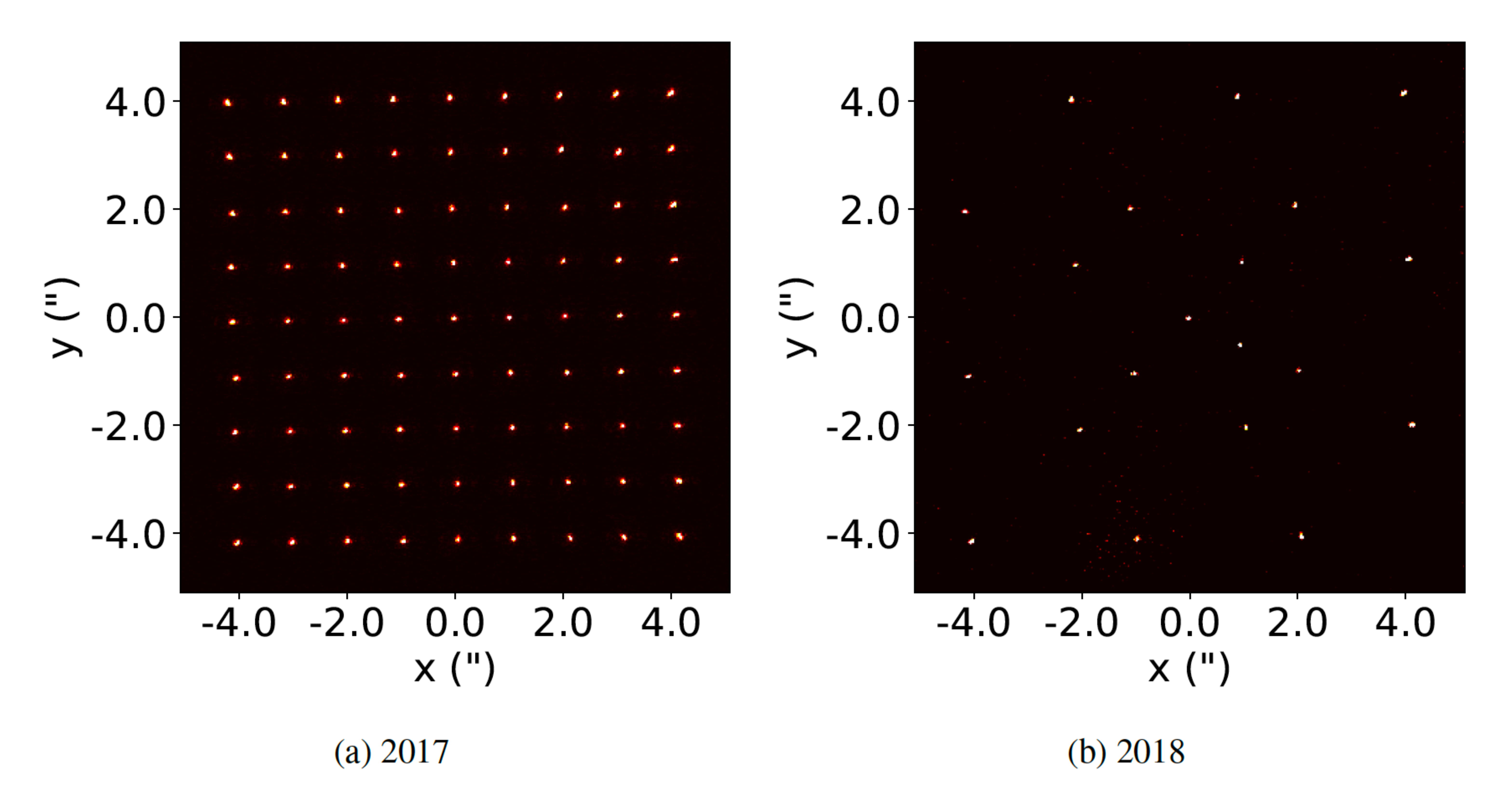}
\caption{Composite images of the Adaptive Optics System calibration fiber in focus, taken by NIRC2 in different years.
The fiber light is only impacted by the optics in the AO system and the NIRC2 instrument.\label{fig:fiber_images}}
\end{figure}

\begin{figure}
\centering
\includegraphics[width=0.9\textwidth]{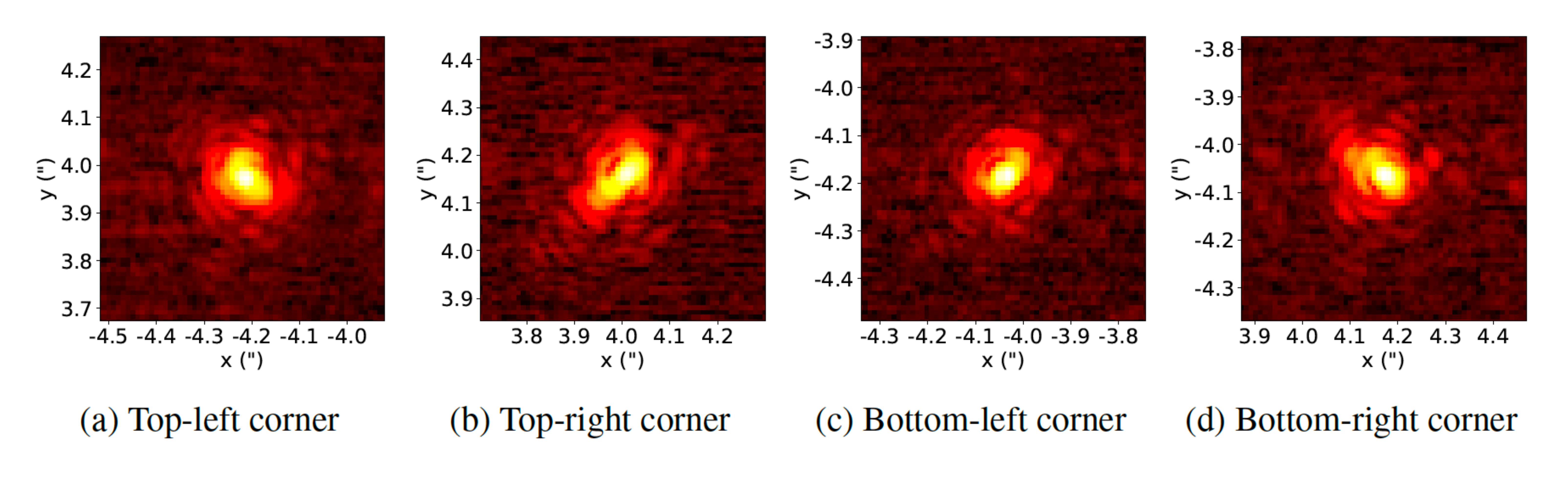}
\caption{Zoomed-in views of the left panel of Figure~\ref{fig:fiber_images} with logarithmic color scale, showing images of the calibration fiber at the four corners.
All four PSFs are a result of the AO+NIRC2 optical system, and are clearly elongated in the direction of the center of the field of view.\label{fig:fiber_images_zoom}}
\end{figure}

In our tests, we have analyzed the composite images of the in focus calibration fiber taken with NIRC2 in 2017 and 2018 (Figure~\ref{fig:fiber_images}).
For the latter, the fiber was positioned in a sparse configuration and not on a grid.
The most noticeable pattern of aberrations is the elongation of the PSF similar to tangential astigmatism, along the direction of the center of the field (Figure~\ref{fig:fiber_images_zoom}).

\begin{figure}
\centering
\includegraphics[width=0.8\textwidth]{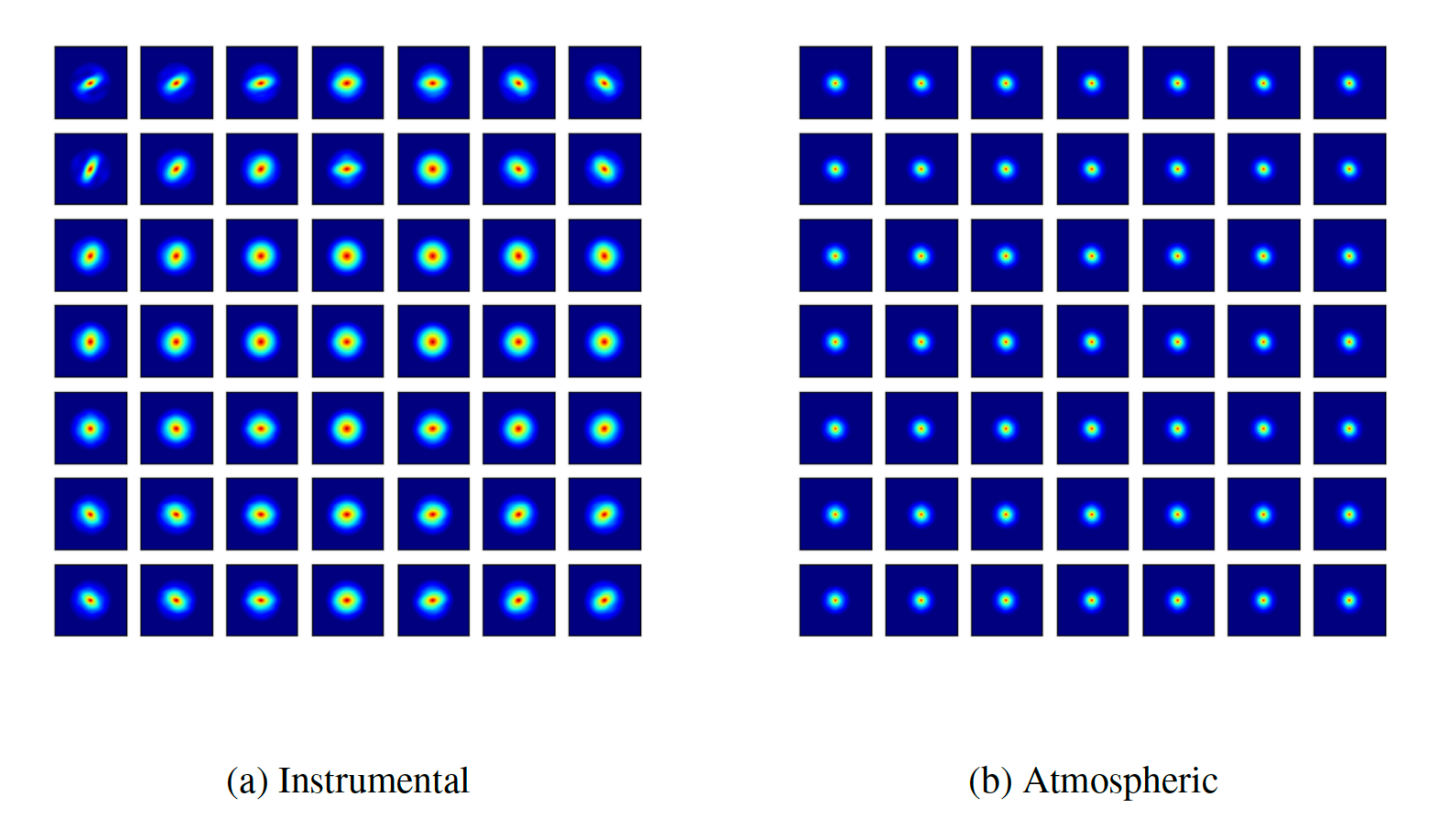}
\caption{MTF sampled at different field positions across the 10\arcsec$\times$10\arcsec\enspace FOV of NIRC2.
The color scale is linear, normalized between 0 (blue) and 1 (red).
The range of angular frequencies of each MTF is $\pm70\,arcsec^{-1}$.
The instrumental MTF was taken in 2017.
The atmospheric MTF is calculated for a turbulence profile with 0.65\arcsec seeing.}\label{fig:otf}
\end{figure}

\begin{figure}
\centering
\includegraphics[width=0.9\textwidth]{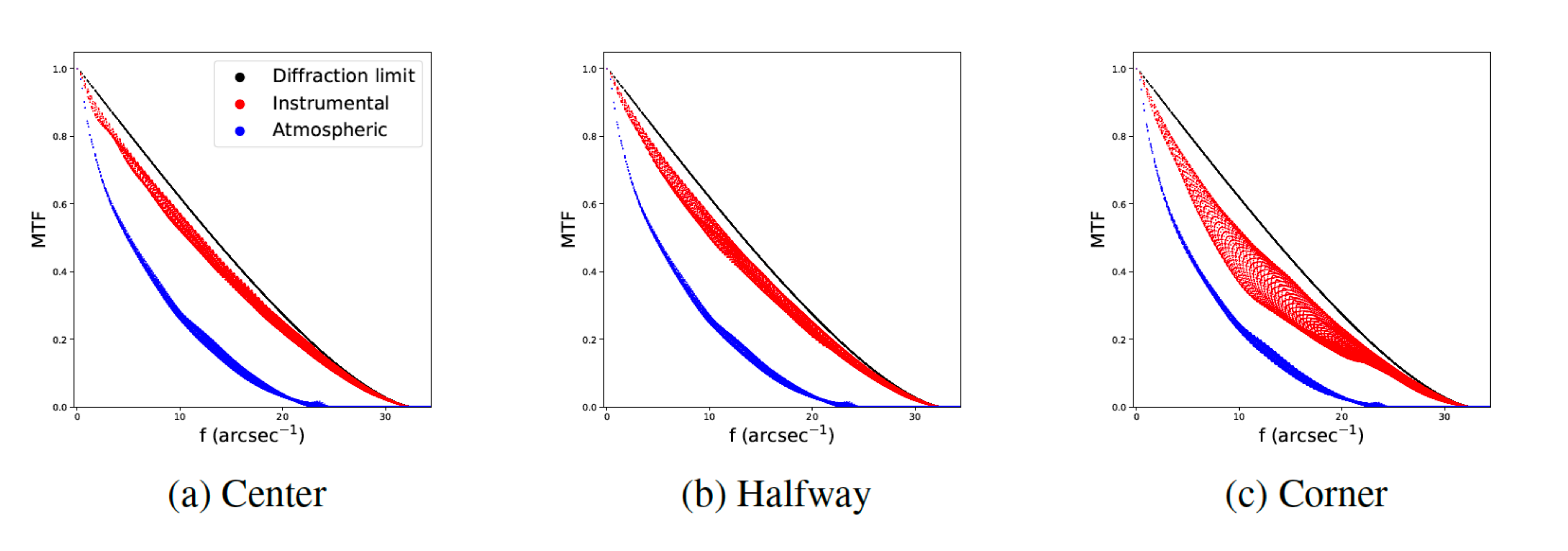}
\caption{MTF at three field positions of the NIRC2 FOV: center, corner, and halfway between the two.
The Instrumental and atmospheric MTF are compared to the diffraction limit of a circular aperture with the diameter of the Keck telescope.}\label{fig:otf_3}
\end{figure}

Instrumental OTFs (Figure~\ref{fig:otf}) are calculated from the 2017 grid of phase maps decomposed using principal component analysis and then projected on a 33$\times$33 grid by cubic convolution interpolation.
The grid is upsampled from the original 9$\times$9 to provide a smoother transition between reconstructed PSFs.
The MTFs shown in Figure~\ref{fig:otf} represent the ability of the optical system to reproduce the contrast of two-dimensional spatial frequencies.
The highest frequencies are at the edge of each panel.
The wider the MTF, the smaller the PSF associated to it, resulting in NIRC2 exposures with higher resolution.
As shown in Figure~\ref{fig:otf_3}, the NIRC2 MTF is close to the diffraction limit in most of the FOV, with considerable deviations only in the corners of the field.

In Section~\ref{sec:calib} we use Figure~\ref{fig:fiber_images} also to test the two AIROPA modes, by fitting the in-focus fiber images with the instrumentation-only PSF.

\subsection{On-sky Images for Testing AIROPA}\label{sec:nirc2img}

\begin{figure}
\centering
\includegraphics[width=0.4\textwidth]{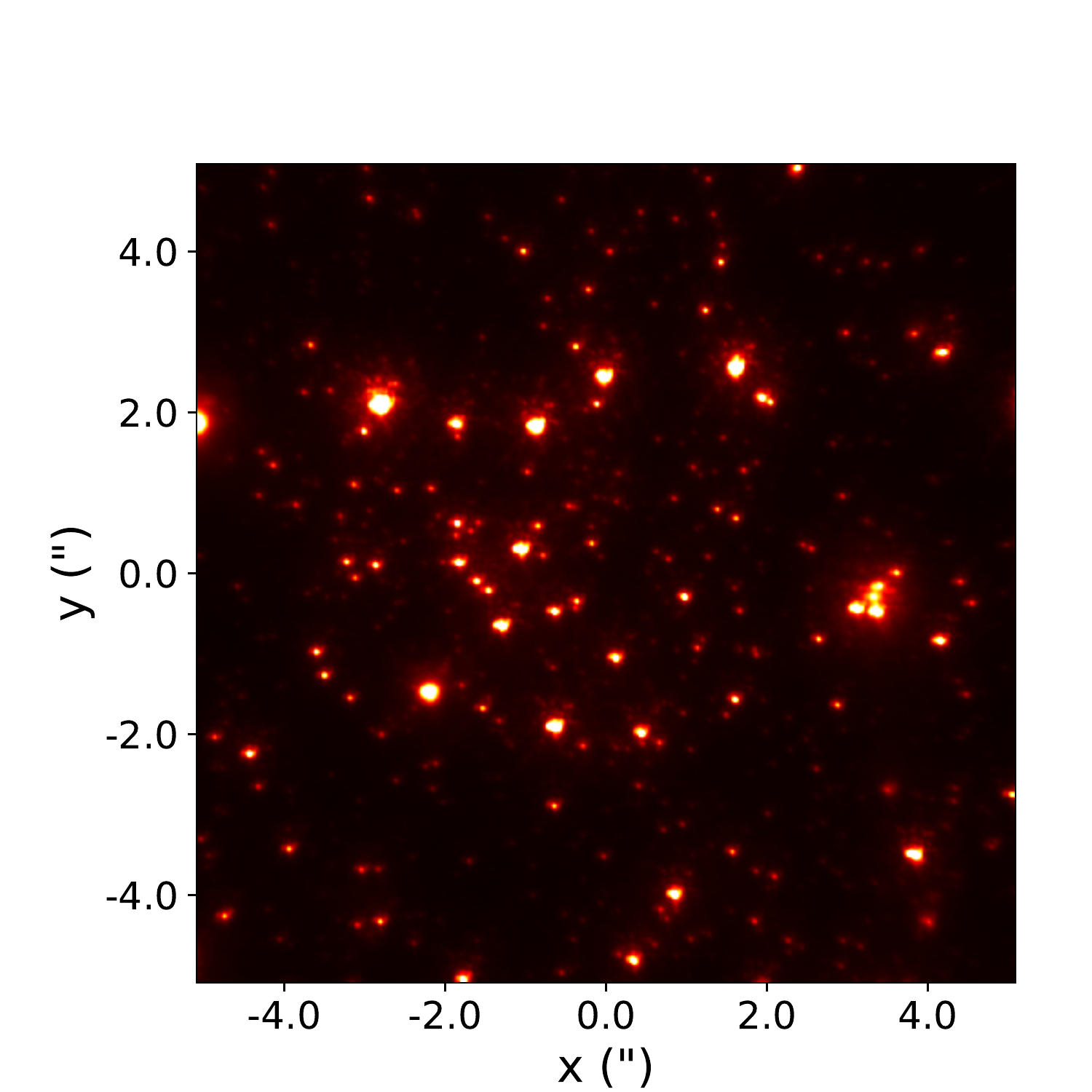}
\caption{A NIRC2 exposure of the GC taken on 2017-08-22 with the Kp filter (FWHM = 71 mas, SR = 0.25).\label{fig:gc_image}}
\end{figure}

\begin{figure}
\centering
\includegraphics[width=0.6\textwidth]{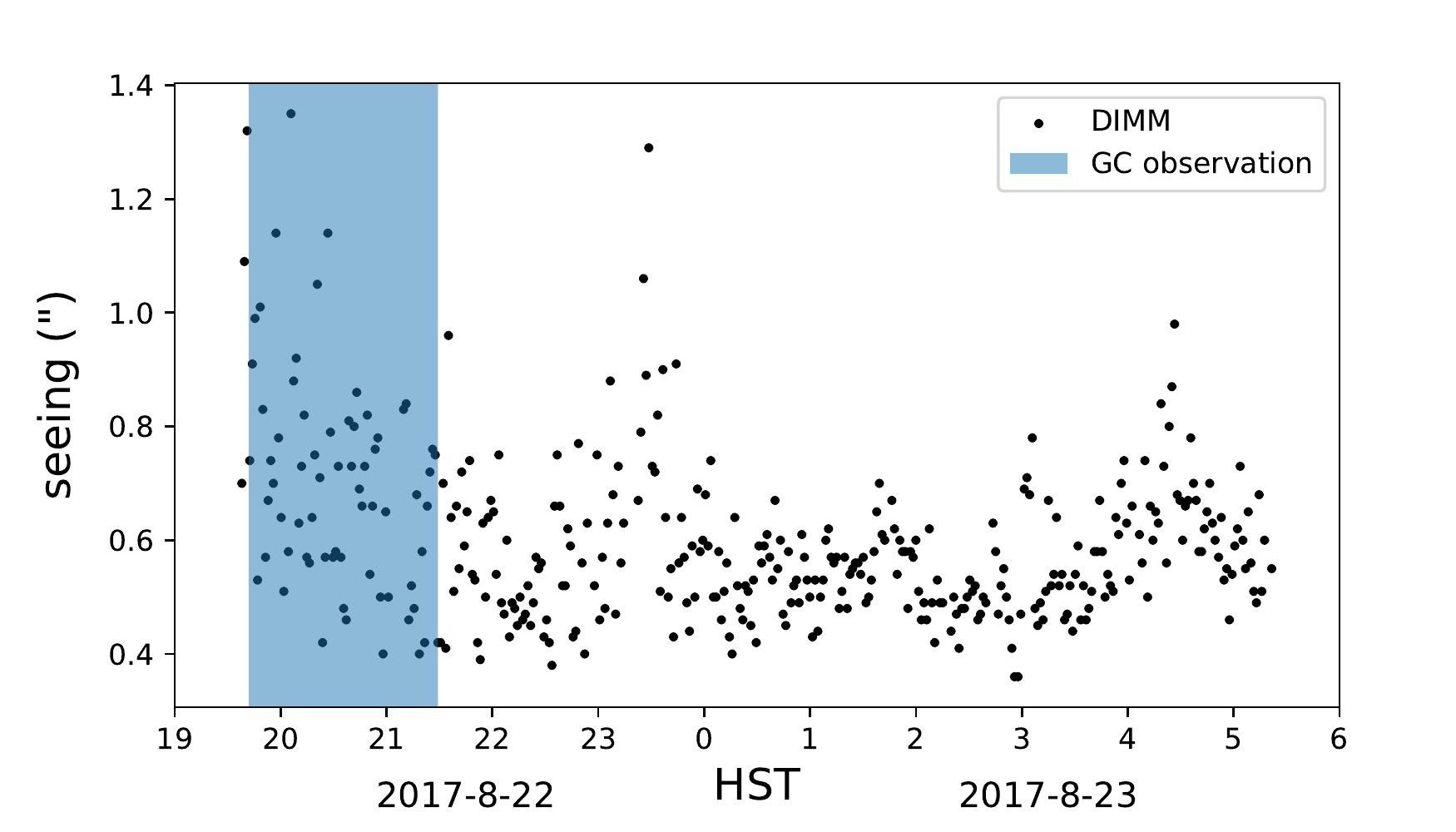}
\caption{Seeing measured by the DIMM instrument (see Section~\ref{sec:atm}) during the night of the GC observation.\label{fig:seeing}}
\end{figure}

For the tests with on-sky data, we have used AIROPA on 116 NIRC2 images of the GC taken with NIRC2 \cite{bib:gautam19,bib:jia19}, each image made of 10 exposures of 2.8 s coadded.
All images have been taken with the Kp filter ($\lambda_{c} =2.1245$ \textmu m) and an example exposure is shown in Figure~\ref{fig:gc_image}.
This is the principal band used for astrometry of the GC at Keck because, compared to shorter wavelengths, the PSF has a higher Strehl ratio (SR), producing a higher signal-to-noise ratio for the stars.
We have chosen to analyze the night between 2017-08-22 and 2017-08-23 HST, because of the low median seeing of 0.69\arcsec during the observation of the GC, despite it being at the beginning of the night, when the seeing is typically stronger (Figure \ref{fig:seeing}).
Since a strong atmospheric turbulence has the effect of smoothing the PSF of a long exposure, the good seeing allows us to evaluate how well the instrumental aberrations are corrected. The GC dataset was reduced using our standard NIRC2 pipeline to remove instrumental signatures, such as flat fielding, bad pixel masking \cite{bib:ghez08,bib:lu09}, and correction for geometric distortions \cite{bib:yelda10,bib:service16}.

\subsection{Star Catalog for Simulated Images}\label{sec:cat}

\begin{figure}
\centering
\includegraphics[width=0.4\textwidth]{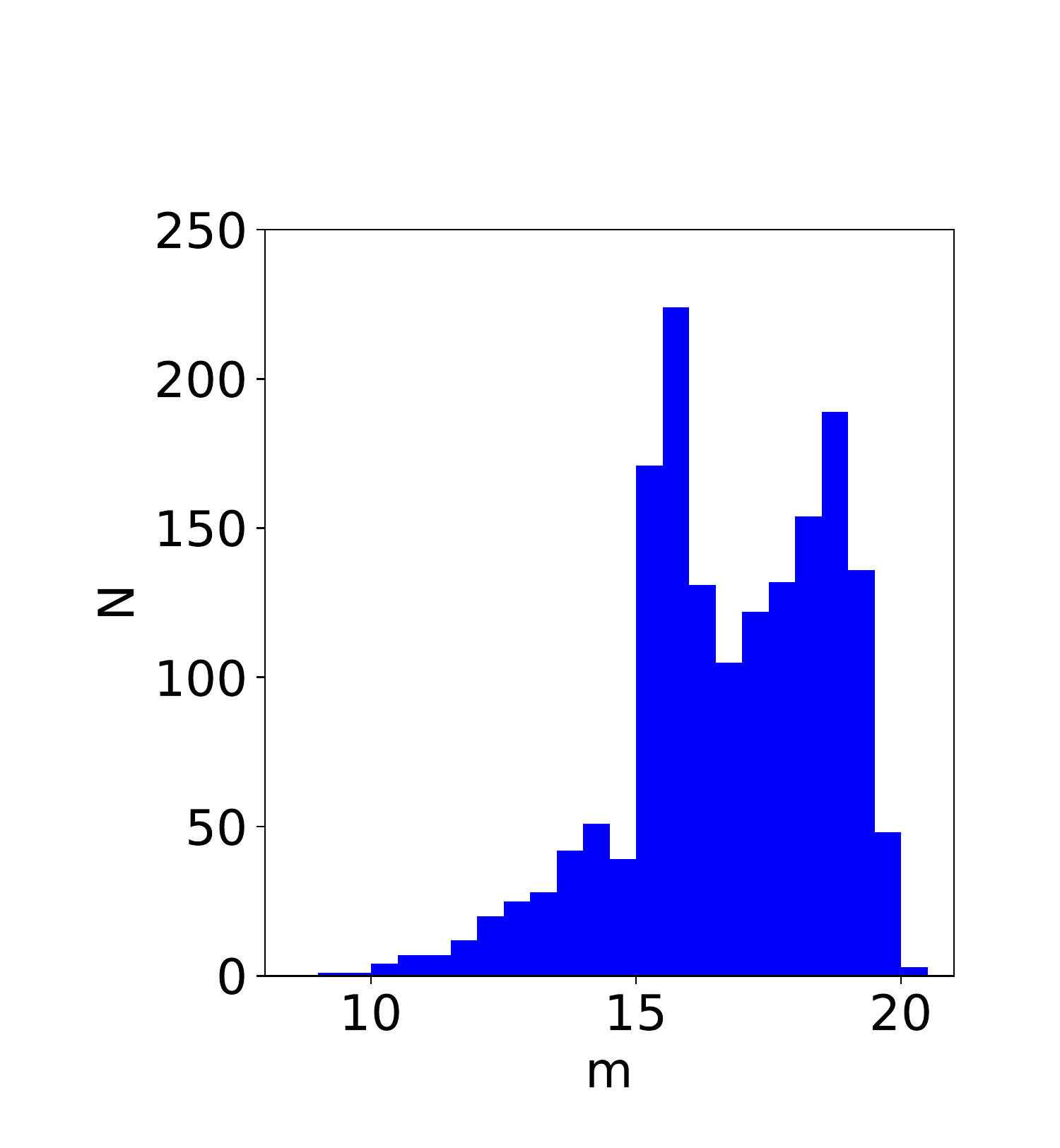}
\caption{Luminosity function of the stars used to simulate the GC.
Magnitudes are in the Kp filter.\label{fig:lum_fun}}
\end{figure}

In addition to the observed images, we have created a simulated NIRC2 image of the GC by using the most recent catalog of known stellar sources derived from observations with Keck \cite{bib:jia19}.
Of these stars, 1652 have their profile completely or partially within the field of view, with the luminosity function shown in Figure~\ref{fig:lum_fun}.

\section{Simulations to Validate AIROPA}\label{sec:sim}

For our initial tests of AIROPA, we have generated NIRC2 images of point sources.
The use of simulated images allows for an accurate measurement of the performance of AIROPA in controlled conditions, where the known input position and luminosity of the sources are compared to the values extracted by AIROPA.
All images have been simulated at $\lambda =2.1245$ \textmu m, the central wavelength of the NIRC2 Kp filter.
All reported magnitude measurements are instrumental, not calibrated to a standard photometric system.
The grid of input PSFs has been calculated from the instrumental phase maps, atmospheric turbulence profile and reference PSF described in Sections~\ref{sec:inst}, \ref{sec:atm}, and \ref{sec:ref}, respectively.

Synthetic images have been produced with the python package \texttt{MAOSI}\footnote{\href{https://github.com/jluastro/maosi}{github.com/jluastro/maosi}} (Make Adaptive Optics Simulated Images), which uses as inputs a catalog of star positions and intensities, a PSF grid, and relevant parameters of the imaging system (read noise, background flux, gain, detector size, pixel scale), together with the exposure time.
Saturation of the detector and non-linearity response are not simulated.

When the simulated image has a spatially variable PSF, the profile of a star at a position in the field of view is computed using a bilinear interpolation of the PSF grid.
Then, the sub-pixel positioning of the PSF onto the detector pixel grid is calculated using a bilinear spline interpolation.
For all simulated images, we have used 200 coadds of 2.8 s exposures.

\subsection{Turbulence}\label{sec:atm}

To model the atmospheric PSF, AIROPA uses information on the seeing and the altitude profile of the turbulence layers $C_{n}^{2}$ \cite{bib:rocca74} taken simultaneously to the science data, or within a few minutes\cite{bib:britton06}.
These parameters are provided, respectively, by the DIMM \cite{bib:martin87, bib:sarazin90} and MASS \cite{bib:kornilov03, bib:tokovinin03} instruments of the Mauna Kea Atmospheric Monitor at the Canada-France-Hawaii Telescope.
We have also employed a single DIMM and MASS measurement from the same night to generate synthetic images for testing, with a seeing of 0.65\arcsec.
The corresponding atmospheric OTF grid simulated by AIROPA is shown in the right panel of Figure~\ref{fig:otf}.
These MTFs are considerably smaller than the instrumental ones in the left panel--as can be seen in Figure~\ref{fig:otf_3}--indicating that the atmosphere is the limiting factor to the spatial resolution of our observations.

\subsection{Reference PSF}\label{sec:ref}

To generate a synthetic PSF and simulate a NIRC2 frame, we need to provide AIROPA with a reasonable model of the PSF containing the aberrations common to all stars in the field of view (see Section~\ref{sec:airopa}).
We have extracted one from a stack of 72 LGS NIRC2 images of a bright star in a binary system with 20\arcsec of separation.
The companion star was used as the NGS, with its magnitude and distance from the science field comparable to the star \mbox{USNO-A2.0 0600.28577051} \mbox{(17$^\mathrm{h}$45$^\mathrm{m}$40$^\mathrm{s}$.72, -29\arcdeg 0\arcmin 11\arcsec .2)} ($m_{R}=13.8$ mag), which is typically employed in GC observations to measure tip-tilt on the Keck AO System \cite{bib:ghez08}.

\subsection{PSF Fitting}\label{sec:fit}

First, the baseline astrometric and photometric performance of AIROPA was determined using simulations of bright sources in a sparse field with a PSF that does not change across the field of view.
The same PSF grids used to simulate the images are then employed for the PSF-fitting.
Because of the optimal PSF-fitting conditions (high signal-to-noise ratio of $\sim 1000$, no crowding, uniform PSF, no PSF extraction), the residuals in position and brightness are interpreted as systematic errors from the underlying StarFinder code and the approximations that it uses.

\begin{figure}
\centering
\includegraphics[width=0.4\textwidth]{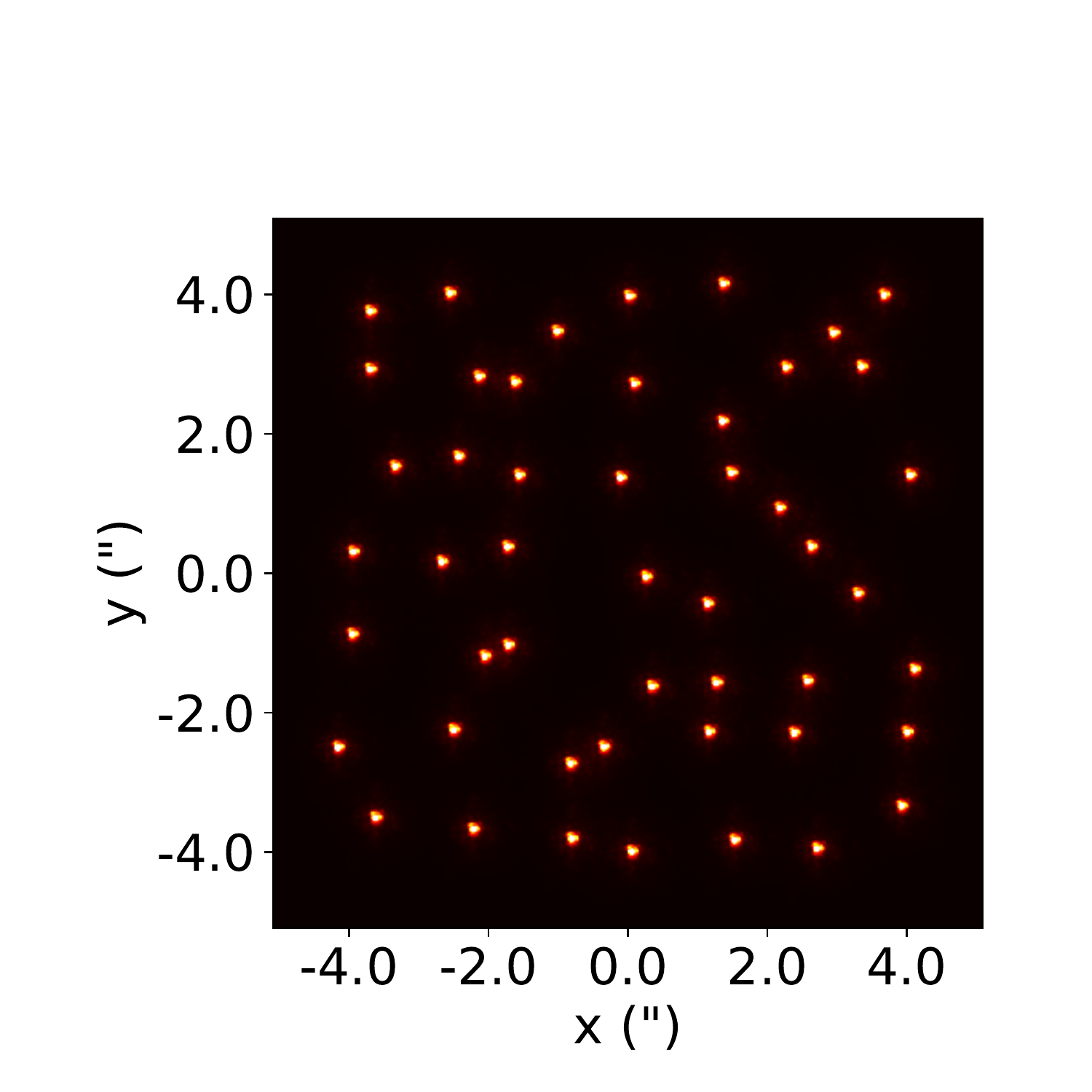}
\caption{Simulation of a NIRC2 image of bright sources with uniform and instrumentation-only PSF.\label{fig:fit_image}}
\end{figure}

The center of the instrumental PSF grid is used as the uniform PSF.
We have produced an image with an irregular 7$\times$7 grid of point sources (Figure~\ref{fig:fit_image}).
The x and y coordinates of the stars have been substantially deviated from a perfect grid by adding a random value with uniform distribution of $\pm0.5$\arcsec.
It was done to avoid a regular pattern of stars that could be mistaken for a feature of the PSF.
If most stars have the halo of other stars in the same relative position, the algorithm extracting the PSF will mistake this additional light as part of the model.

\begin{figure}
\centering
\includegraphics[width=0.7\textwidth]{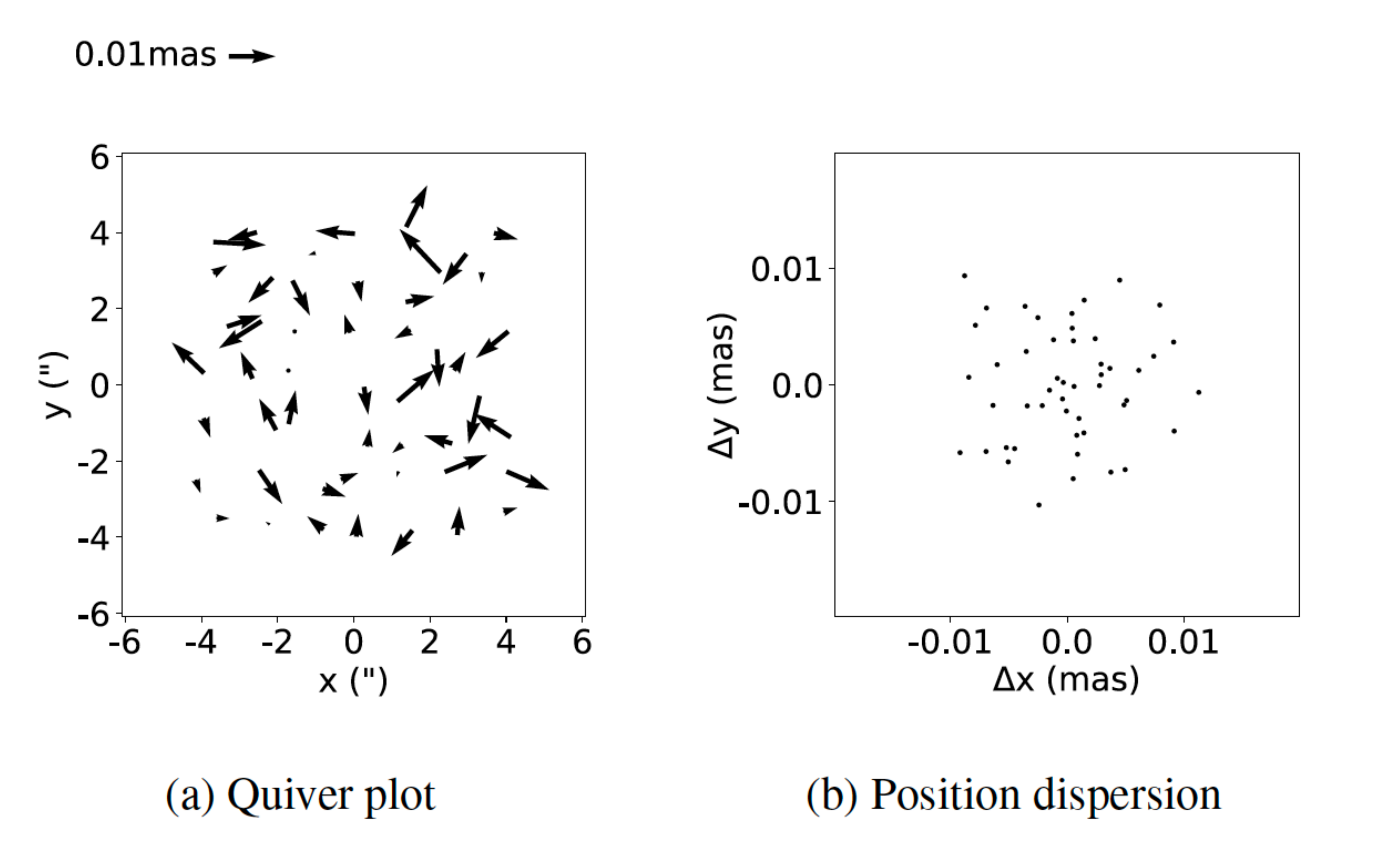}
\caption{Astrometric residuals of simulated bright sources created with a uniform PSF and extracted with AIROPA in \texttt{single-PSF} mode.
The input PSF is from an observed on-axis fiber source image.
\label{fig:fit_res}}
\end{figure}

The image was then analyzed using the \texttt{single-PSF} mode of AIROPA, using all 49 sources as PSF stars.
The astrometric residuals are defined as the difference between the input and output positions in two directions: $\Delta x$, $\Delta y$.
The average astrometric residuals, when injecting with a single PSF and recovering with a single PSF, are \mbox{$\Delta r=\sqrt{\Delta x^{2}+\Delta y^{2}}$} is $6.2\cdot10^{-3}$ mas.
Differences between the input and output positions are shown in Figure~\ref{fig:fit_res}.
The extracted photometry too is extremely consistent with the input values, with the average of the absolute photometric residuals $\left|\Delta m_{Kp}\right|$ at $2\cdot10^{-4}$ mag.
$\Delta r$ and $\Delta m_{Kp}$ have very small values because we have removed almost completely random noises and systematic errors from the measurements, leaving only the pixel sampling error.
They represent therefore the best astrometric and photometric precision possible in ideal conditions with AIROPA.

\subsection{PSF Spatial Variability}\label{sec:psf_var}

Accurate PSF-fitting of images with an AO-corrected extended field of views requires the correct mapping of the spatial variations of the PSF.
To characterize the ability of AIROPA to reconstruct and fit variable PSFs, we have generated two images similarly to the one in Figure~\ref{fig:fit_image}, except for the use of a variable PSF.
In the PSF-reconstruction algorithm of AIROPA, the instrumental and atmospheric OTF ratio grids are calculated independently and then are combined.
The first of the synthetic frames is built from the instrumental OTF only, while the other simulates also atmospheric aberrations.
The atmospheric parameters used were derived from the data described in Section~\ref{sec:atm}.
The LGS is simulated at the center of the frame and the T/T star is on the pixel position \mbox{(-391, 2463)}.
The zenith angle is 45\arcdeg.
This configuration reproduces typical NIRC2 observations of the GC, with the star \mbox{USNO-A2.0 0600.28577051} for tip-tilt sensing.

\begin{figure}
\centering
\includegraphics[width=0.9\textwidth]{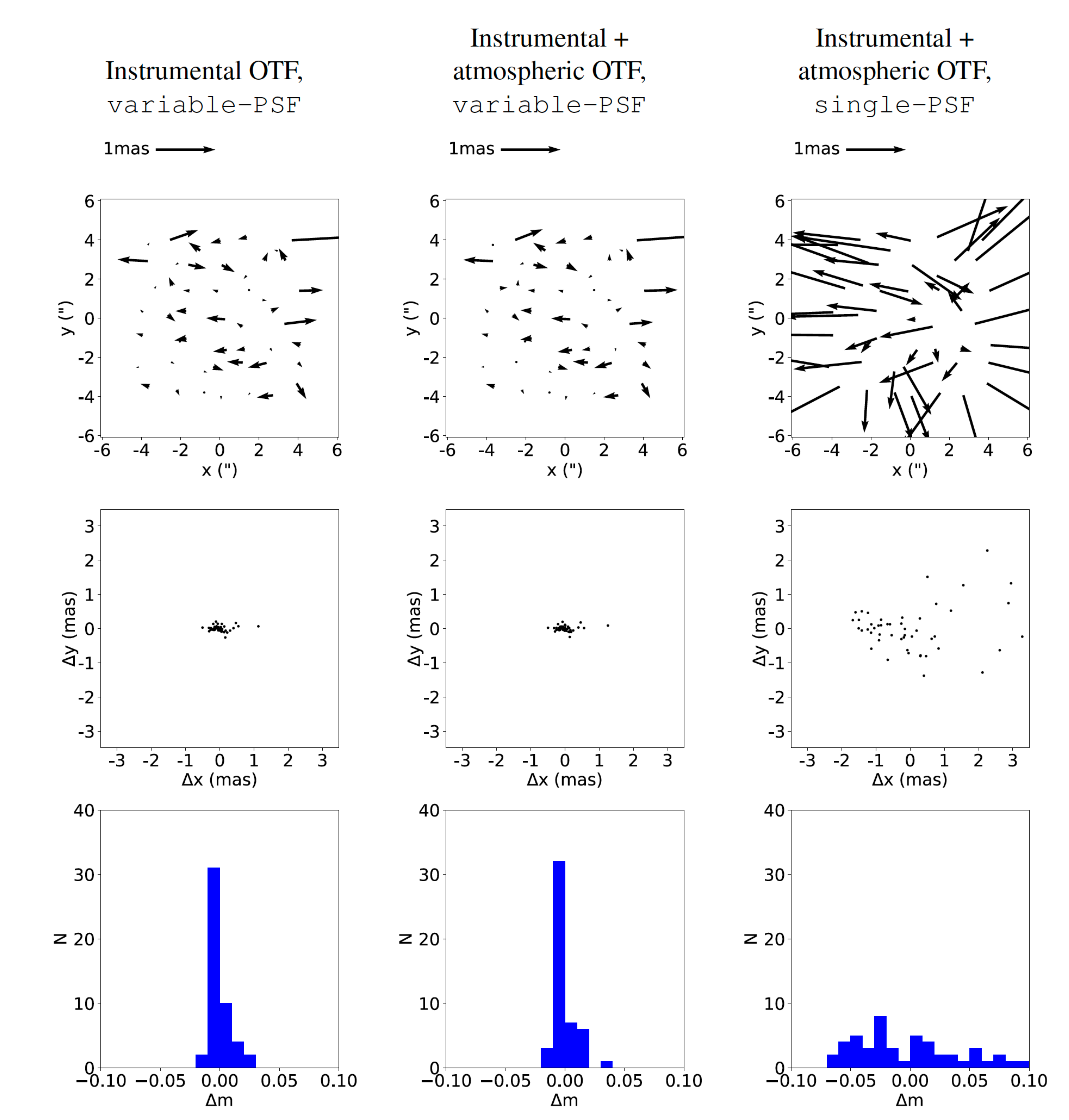}
\caption{Astrometric (first and second row) and photometric (third row) residuals in Kp of simulated bright sources with non-uniform PSF.
The use of a spatially variable PSF (first and second column) produces better residuals than a uniform one (right column).\label{fig:var_res}}
\end{figure}

When using AIROPA in \texttt{variable-PSF} mode with the instrumentation-only image, the average astrometric and photometric residuals are $1.9\cdot10^{-1}$ mas and $5.5\cdot10^{-3}$ mag, respectively.
When the atmospheric OTF is also included, the two values are $\Delta r=1.9\cdot10^{-1}$ mas and $\left|\Delta m_{Kp}\right|=6.1\cdot10^{-3}$ mag.
The residuals are comparable between the two tests (left and central columns of Figure~\ref{fig:var_res}), but are significantly larger than with the a constant-PSF image in Section~\ref{sec:fit}.

\begin{figure}
\centering
\includegraphics[width=0.7\textwidth]{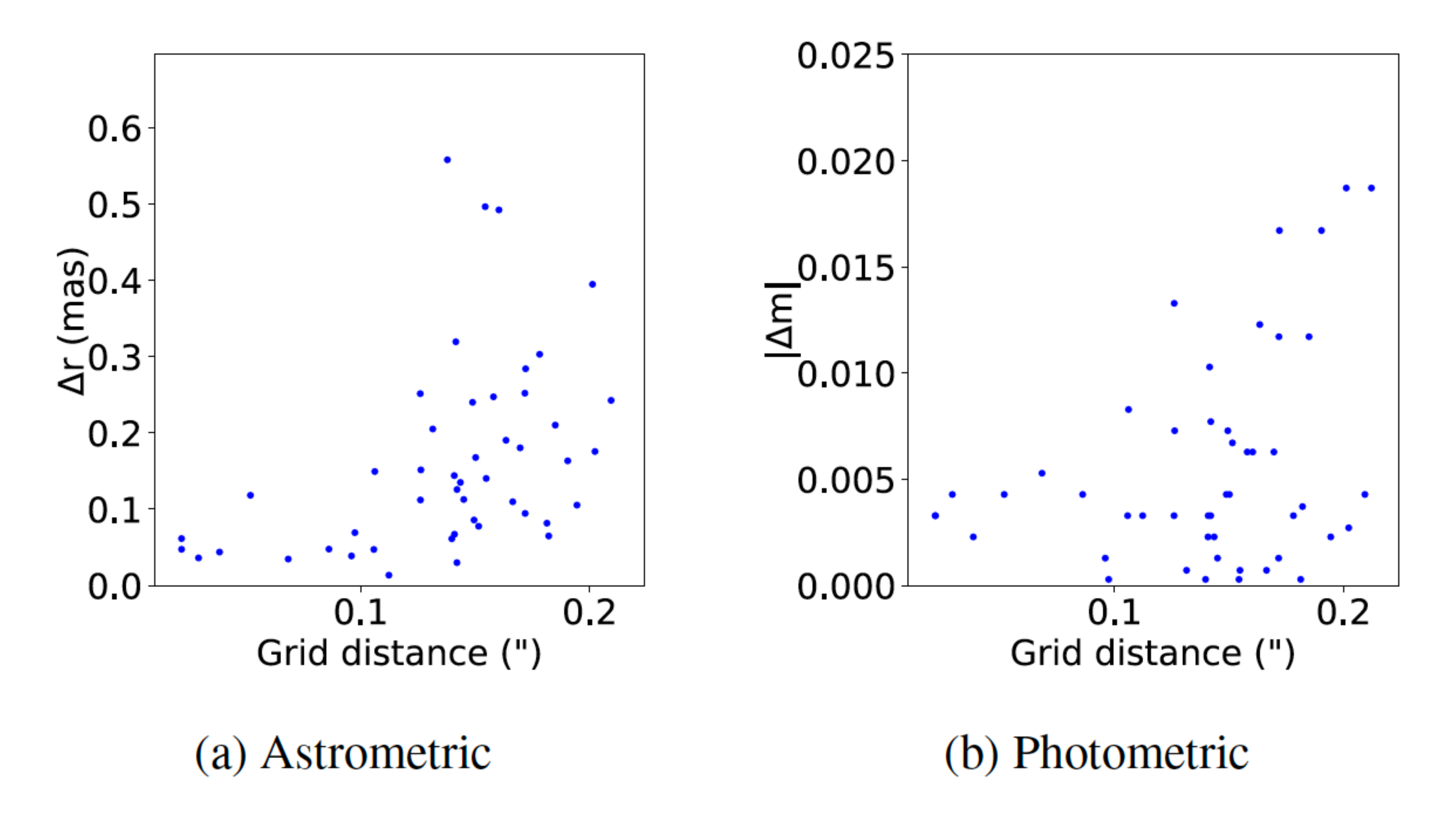}
\caption{Residuals of simulated bright sources with non-uniform, instrumental and atmospheric PSF, after using \texttt{variable-PSF} mode, as a function of their distance from the center of the corresponding PSF grid cell.
Magnitude errors are in the Kp filter.\label{fig:var_res_phase}}
\end{figure}

To understand the deterioration of the results when dealing with a variable PSF, we have to consider how the PSF model is used differently for simulating the image and fitting it.
While MAOSI smooths the PSF between positions of the grid to reproduce its gradual variation as in real instruments (see Section~\ref{sec:sim}), StarFinder and several other PSF-fitting software programs use the nearest neighbor approach.
The discrepancy between the PSF used to generate a star and to fit it, causes a loss in accuracy of more than an order of magnitude in both astrometry and photometry.
This effect can be better appreciated by looking at the increment in residuals as a function of the distance between a star and the center of the grid cell of its PSF (Figure~\ref{fig:var_res_phase}).
While a finer grid would reduce these errors, the maximum practical grid resolution is determined by the density of the instrumental phase map (Section~\ref{sec:inst}).

\begin{figure}
\centering
\includegraphics[width=0.7\textwidth]{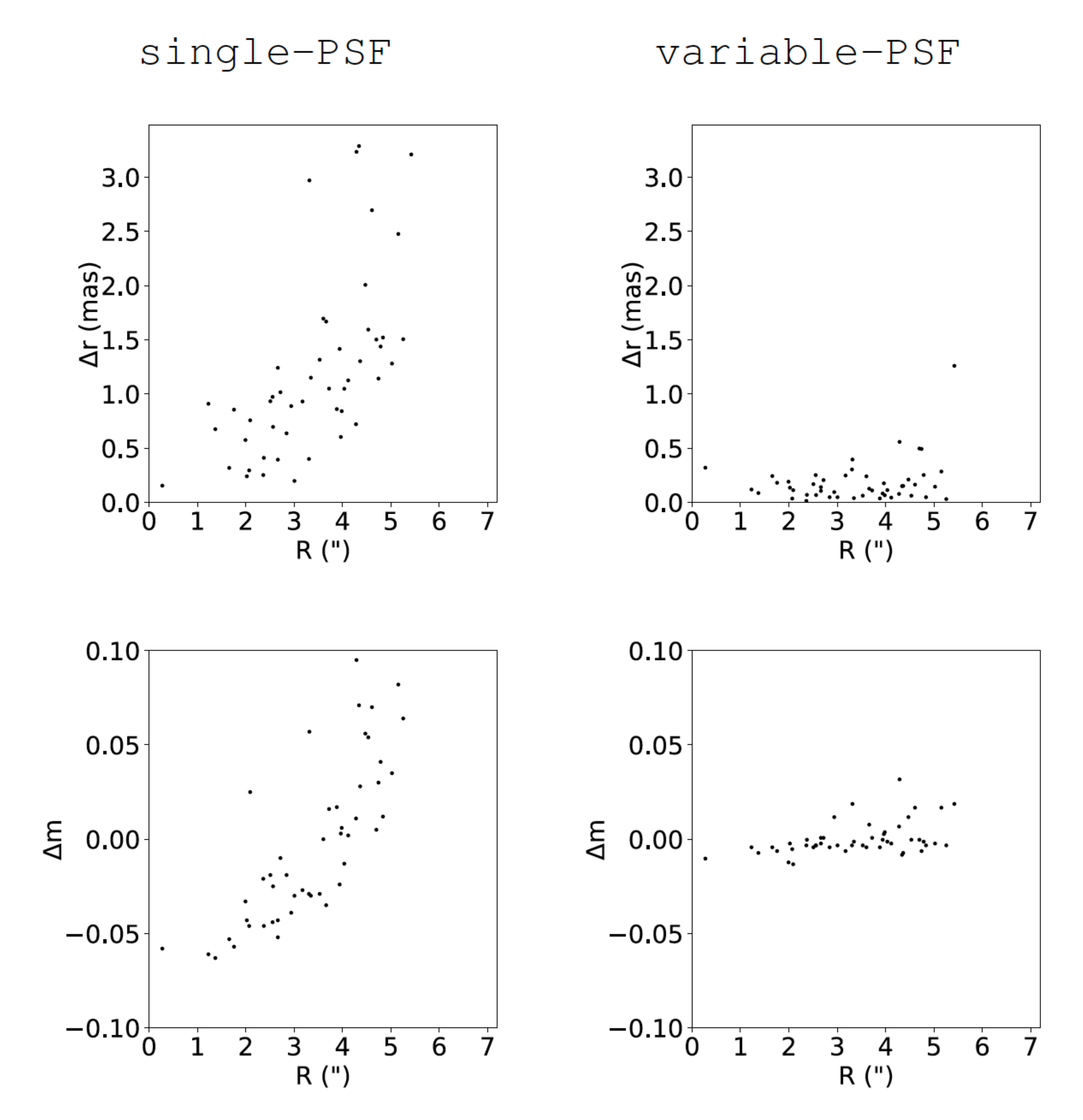}
\caption{Astrometric (top row) and photometric (bottom row) residuals in Kp of a grid of stars with instrumental and atmospheric variable PSF, as function of distance from the center of the image.\label{fig:var_rad}}
\end{figure}

We then run AIROPA in \texttt{single-PSF} mode on the \texttt{variable-PSF} simulated images to replicate the current on-sky analysis. This allows us to evaluate how large of an error is produced by ignoring the PSF spatial variability. The results can be seen in the right column of plots in Figure~\ref{fig:var_res}.
The astrometric and photometric residuals increase to $1.2$ mas and $3.9\cdot10^{-2}$ mag, respectively.
The quiver plot for the \texttt{single-PSF} mode in Figure~\ref{fig:var_res} shows a dominant radial pattern for the astrometric residuals.
The correlation between the \texttt{single-PSF} residuals and the distance of the stars from the center of the field of view is more obvious from the left column of panels in Figure~\ref{fig:var_rad}).
AIROPA's \texttt{variable-PSF} mode ability of making residuals more uniform is shown on the right column.

\subsection{GC Simulations}\label{sec:psf_gal}

\begin{figure}
\centering
\includegraphics[width=0.9\textwidth]{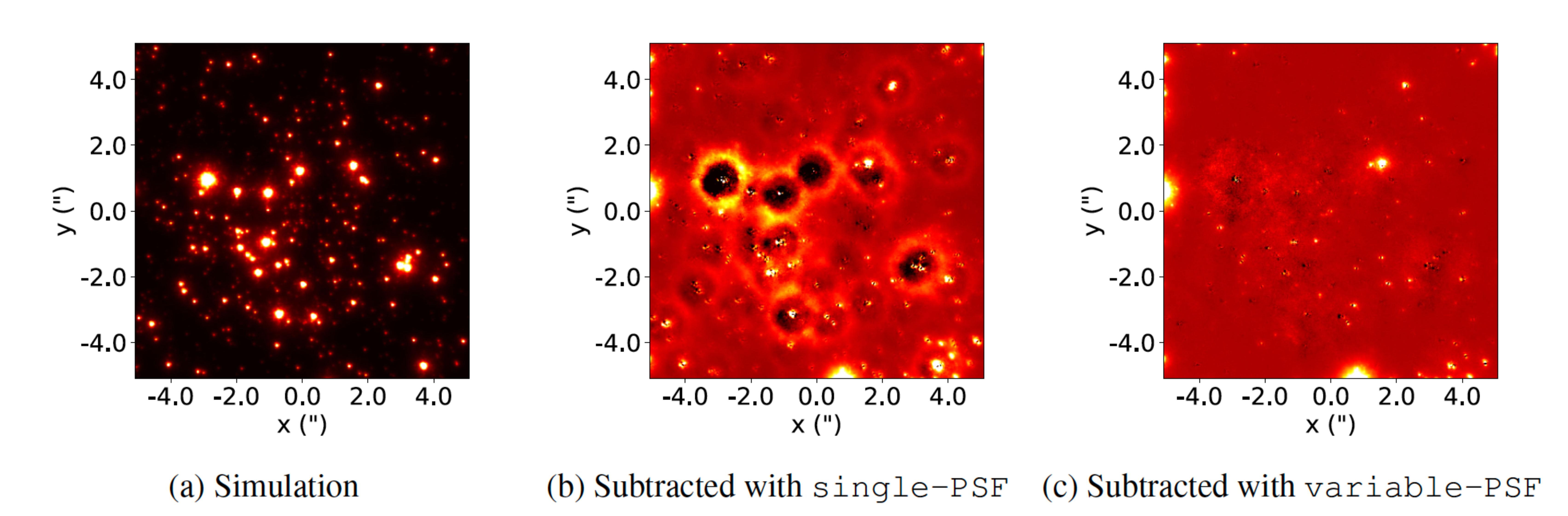}
\caption{Simulated and residual images of the GC.
The same color scale is used for the \texttt{single-PSF} and \texttt{variable-PSF} subtracted images.\label{fig:gc_sim}}
\end{figure}

A more faithful simulation of the systematic errors that AIROPA could face with on-sky images requires sources with a realistic luminosity function and density.
We have chosen to use the GC as a representative target for observations with NIRC2 because of its crowding and the wide range of magnitudes.
The purpose of this test is to assess the combined accuracy of PSF reconstruction and PSF fitting algorithms in AIROPA.
The simulation of the central 10\arcsec of the Galactic SMBH (Figure~\ref{fig:gc_sim}) uses the same conditions (zenith angle, DIMM and MASS profile data, LGS and T/T star positions) of the tests in Section~\ref{sec:psf_var}.

\begin{figure}
\centering
\includegraphics[width=0.9\textwidth]{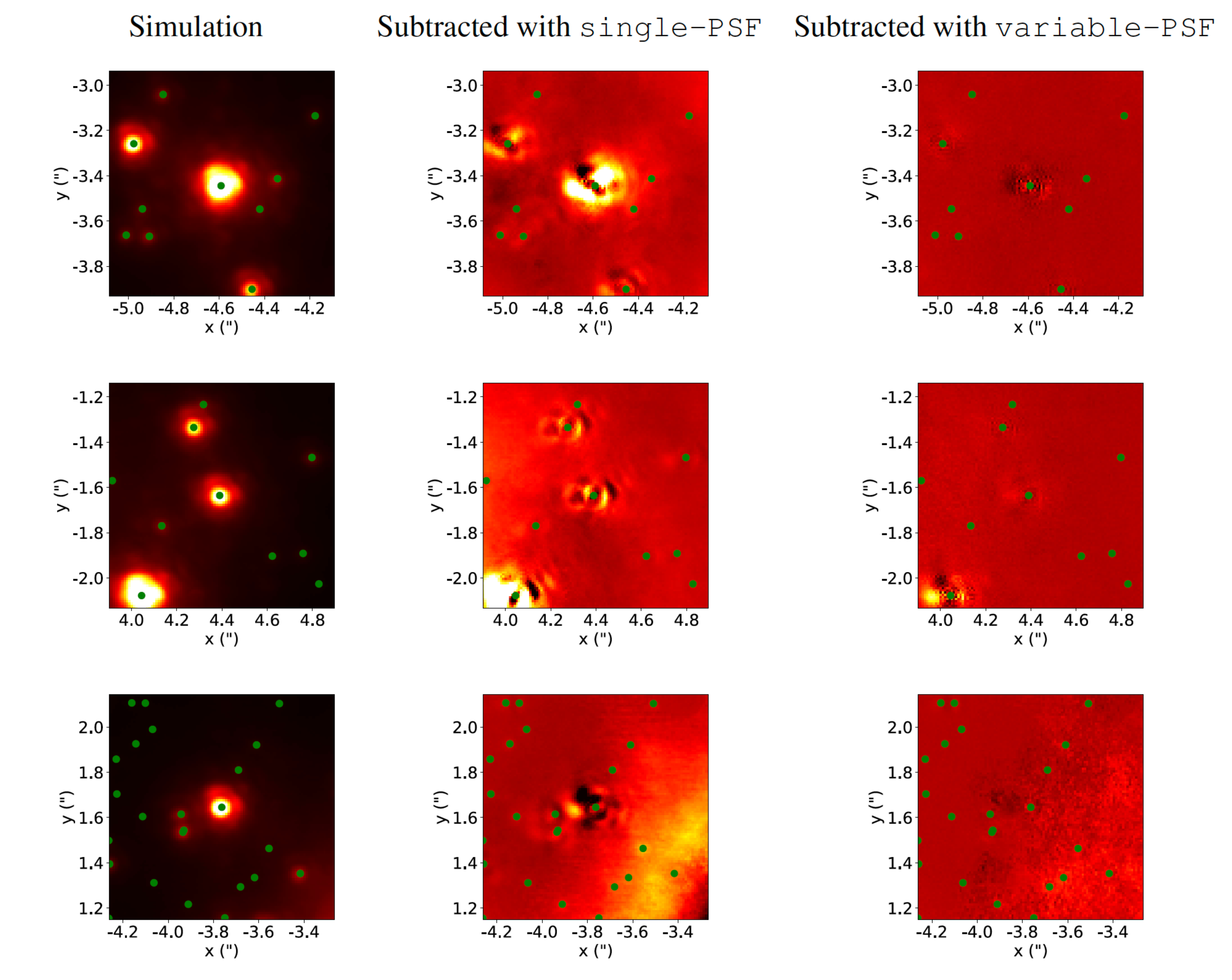}
\caption{Simulated and residual images of several stars in the GC.
The \texttt{single-PSF} and \texttt{variable-PSF} subtracted images have the same color scale.
The green dots show the location where stars were planted.\label{fig:gc_sim_res_stars}}
\end{figure}

As in the previous Section, the \texttt{variable-PSF} mode provides a clear improvement on the astrometry and photometry over the \texttt{single-PSF} mode.
The image after subtracting the stars with a variable PSF (right panel of Figure~\ref{fig:gc_sim}) has much cleaner residuals than the one using a constant PSF (central panel of Figure~\ref{fig:gc_sim}), both in the core and in the halo of the PSF, as can be observed in the close-ups in Figure~\ref{fig:gc_sim_res_stars}.

\begin{figure}
\centering
\includegraphics[width=0.7\textwidth]{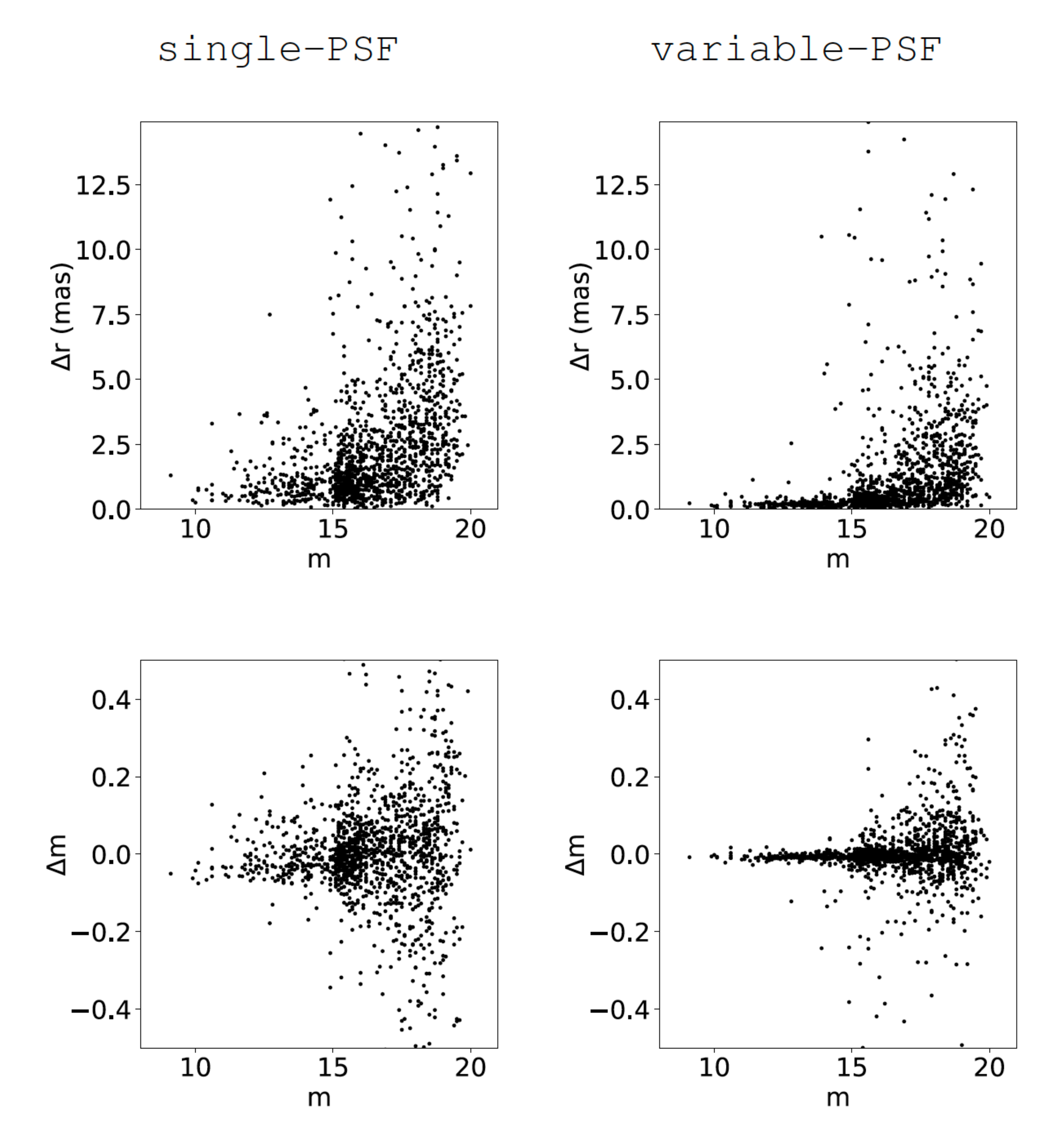}
\caption{Astrometric (first row) and photometric (second row) residuals in Kp with AIROPA on a simulated GC image, as a function of magnitude.\label{fig:gc_sim_mag}}
\end{figure}

To quantify the performance of the two modes, we have considered the average residuals of only the brightest stars that have instrumental magnitude $m_{Kp}\leq 14$.
Up to this magnitude, PSF-fitting residuals are mostly affected by systematic errors in the model of the PSF rather than by noise (Figure~\ref{fig:gc_sim_mag}).

\begin{figure}
\centering
\includegraphics[width=0.7\textwidth]{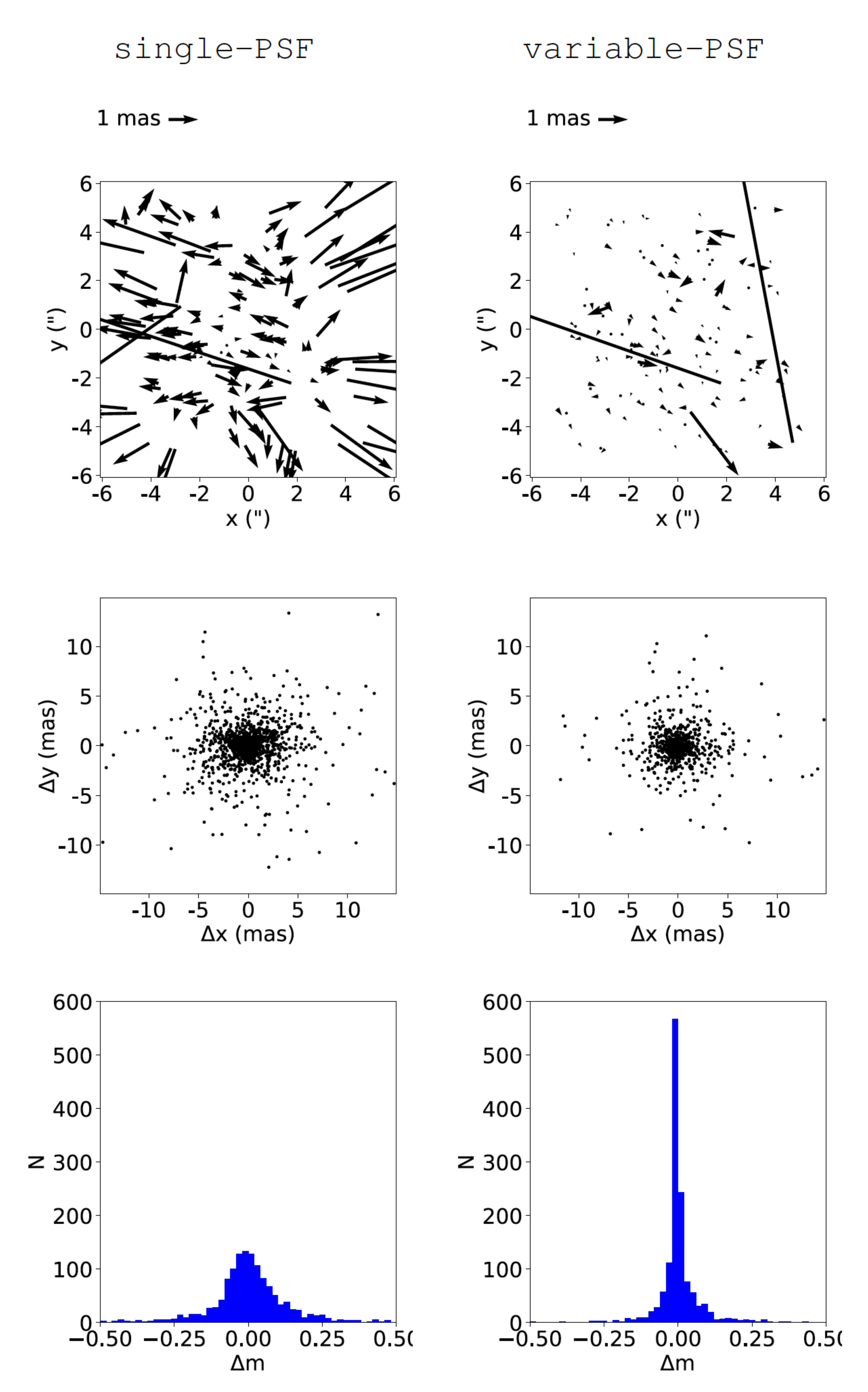}
\caption{Astrometric (first and second row) and photometric (third row) residuals in Kp with AIROPA on a simulated GC image.
The quiver plots show only bright stars ($m_{Kp}\leq 14$).\label{fig:gc_sim_res}}
\end{figure}

For bright stars, astrometric residuals are greatly reduced from $8.3\cdot10^{-1}$ mas to $1.7\cdot10^{-1}$ mas, and the same is true for photometric residuals, lowered from $3.7\cdot10^{-2}$ mag to $0.6\cdot10^{-2}$ mag (Figure~\ref{fig:gc_sim_res}).
The values for the \texttt{variable-PSF} mode are in line with those in the test with the sparse field (Section~\ref{sec:psf_var}), indicating that a level of crowding like the one in the GC is efficiently managed by the PSF-fitting algorithm.

\begin{figure}
\centering
\includegraphics[width=0.7\textwidth]{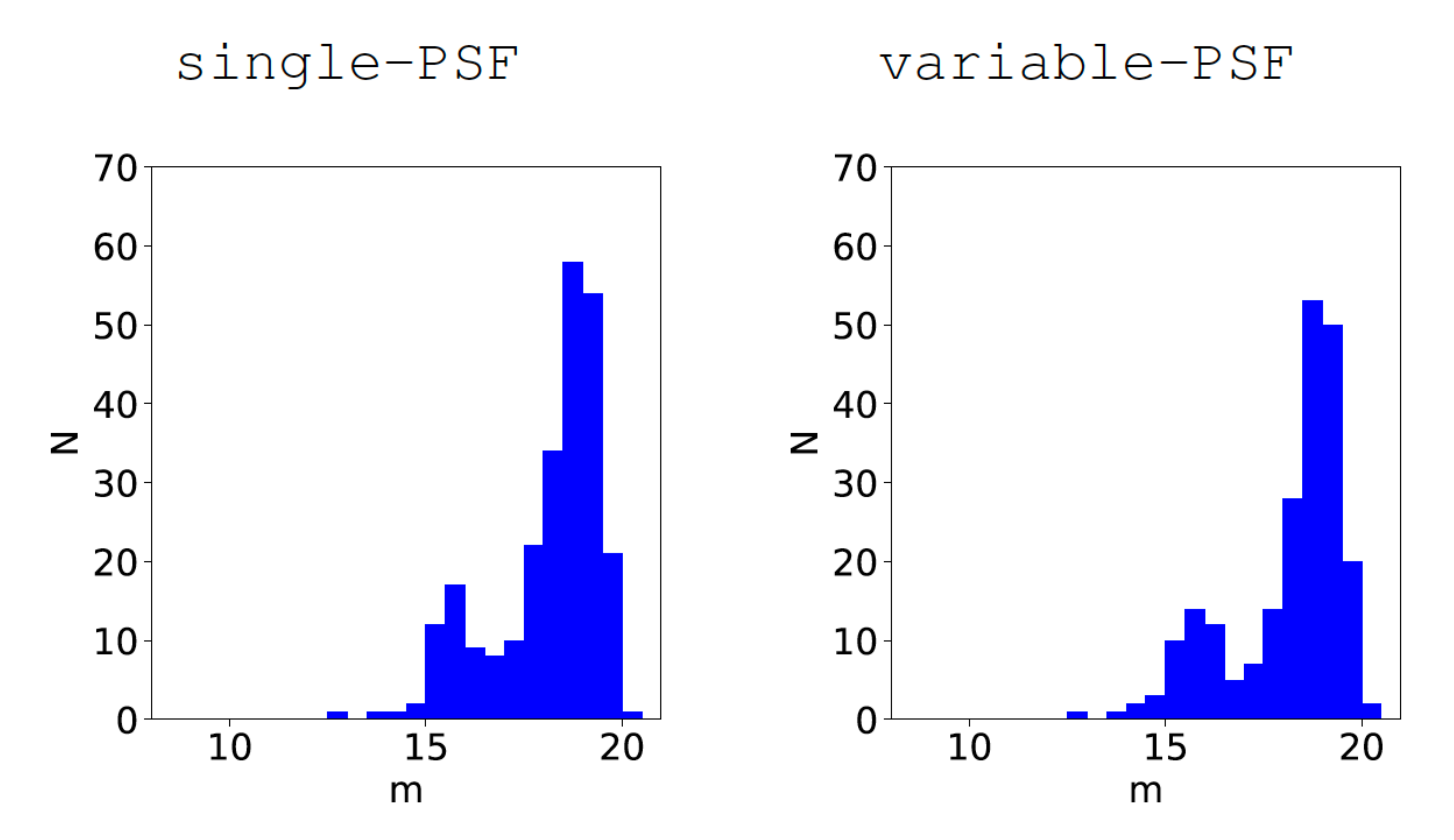}
\caption{Kp luminosity function of the stars planted, but not detected by PSF-fitting in the simulated image of the GC.\label{fig:gc_sim_miss}}
\end{figure}

A test image crowded with stars with a wide range of magnitudes is useful also to investigate the ability of AIROPA to find all and only the stars inserted in the simulation.
Fewer input stars should be missed (false negatives) and fewer fake stars should be detected (false positives) with the \texttt{variable-PSF} mode if it is as accurate as expected.
The \texttt{single-PSF} and \texttt{variable-PSF} modes in AIROPA miss a similar number of sources, 251 versus 222 respectively (Figure~\ref{fig:gc_sim_miss}), with a slight improvement when using the reconstructed PSF.
The detection threshold in Starfinder is the minimum correlation between the image of a star and the PSF fitted to it \cite{bib:diolaiti00}.
For both modes in AIROPA, we have used a threshold of 0.8.

\begin{figure}
\centering
\includegraphics[width=0.7\textwidth]{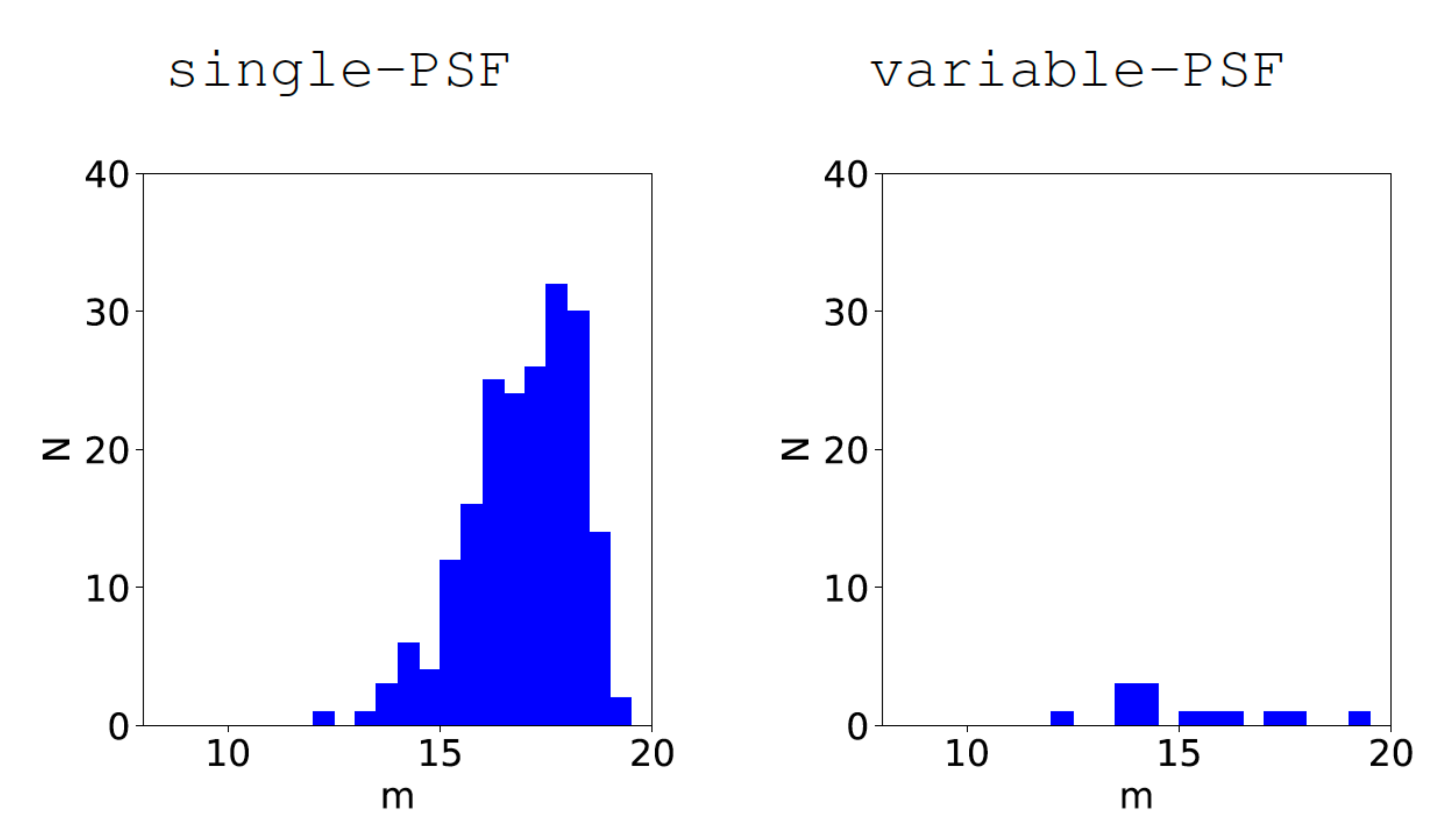}
\caption{Kp luminosity function of the stars detected, but not planted, in the GC simulation.\label{fig:gc_sim_bad}}
\end{figure}

A random peak in the noise or a speckle of a bright source can be interpreted incorrectly as a real star by the PSF-fitting software.
Using \texttt{variable-PSF} greatly reduces the number of erroneous detection, from 197 to 13 (Figure~\ref{fig:gc_sim_bad}).
This improvement is caused by the ability of the reconstructed PSF to reproduce better the images of stars, resulting in less features of their profiles left unfitted and misinterpreted as separate sources.
The small number of wrong detections in the \texttt{variable-PSF} case are mostly noise spikes of the background, unaffected by the choice of PSF-reconstruction algorithm.

A useful metric to measure the residuals of the PSF-fitting is the fraction of variance unexplained (FVU) \cite{bib:do18,bib:ciurlo21}, defined as the ratio between the variance of the residuals and the variance of the image itself:
\begin{equation}
FVU=\frac{\sigma^{2}_{res}}{\sigma^{2}_{img}}
\end{equation}
The fitting errors of bright stars are impacted minimally by noise, and are mostly caused by systematic errors of the PSF.
Therefore, the FVU of bright stars is a better indicator of the ability to reconstruct the PSF than in faint stars.
Smaller FVU values indicate a better fit.
Using the \texttt{single-PSF} mode on the GC simulation, we measure a median FVU of $5.7\cdot10^{-4}$.
By fitting the same stars with the reconstructed PSF, the median value decreases by an order of magnitude, $6.2\cdot10^{-5}$.

\begin{figure}
\centering
\includegraphics[width=0.7\textwidth]{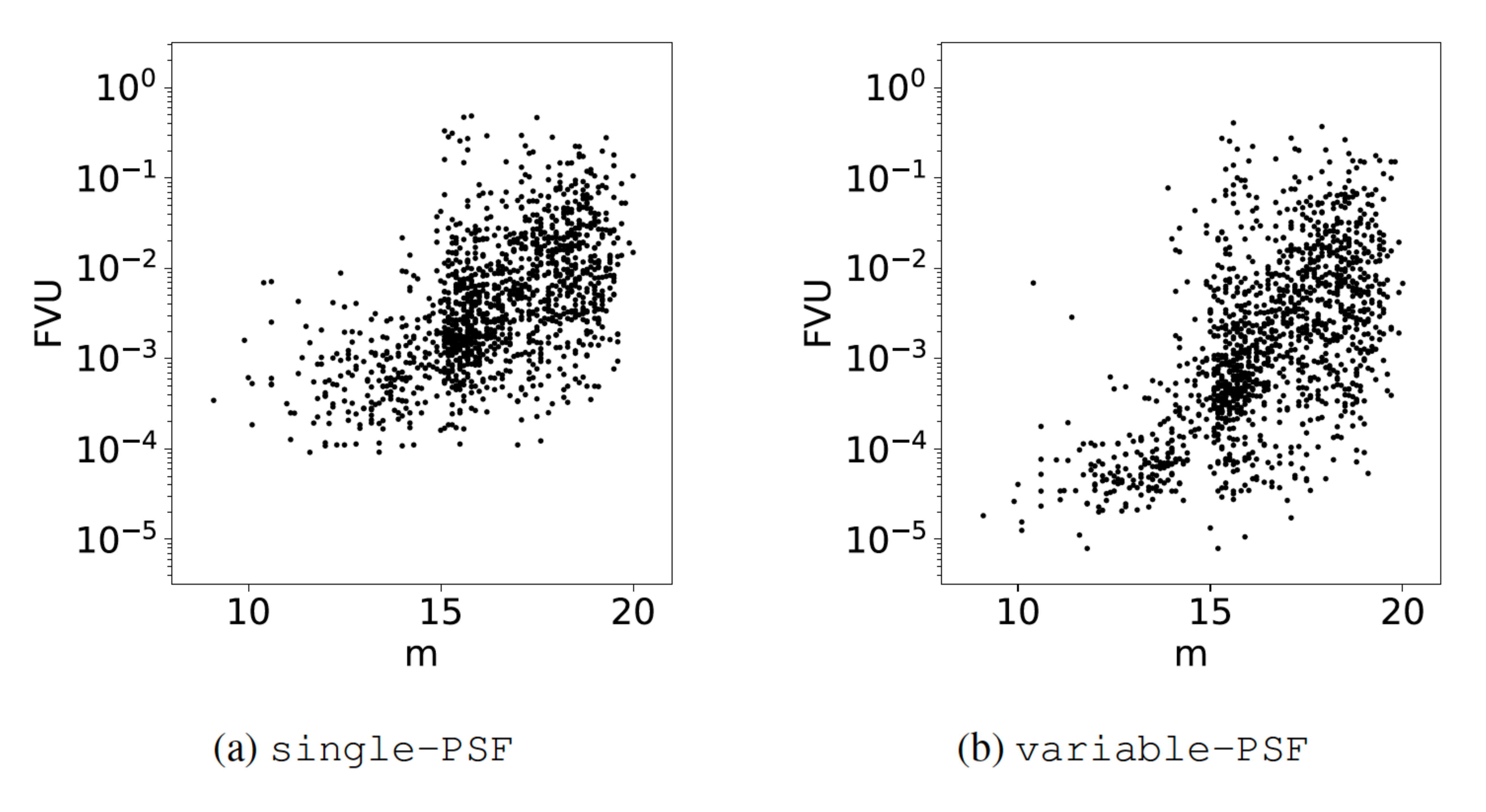}
\caption{FVU of the stars in the GC simulation.
Magnitudes are in the Kp filter.\label{fig:gc_sim_fvu}}
\end{figure}

\begin{figure}
\centering
\includegraphics[width=0.7\textwidth]{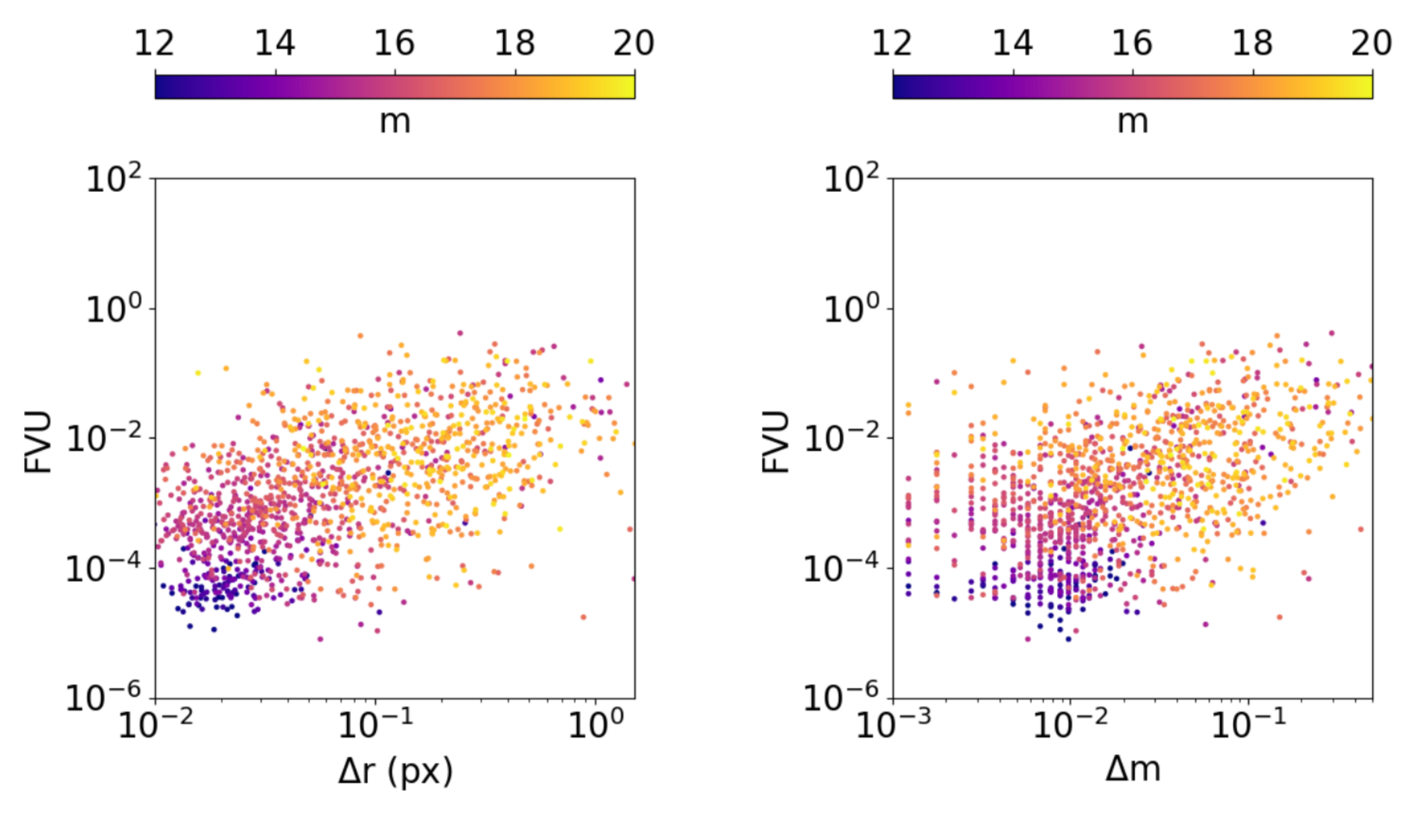}
\caption{FVU of the stars in the GC simulation, as a function of astrometric and photometric errors.
Data points are coloured by instrumental magnitudes in the Kp filter.\label{fig:gc_sim_fvu_drdm}}
\end{figure}

The relation between FVU and astrometric or photometric errors is complex, depending on multiple factors such as the specific shape of the PSF, or the crowding level.
Nevertheless, for NIRC2 observations of the GC, we can estimate it from Figure~\ref{fig:gc_sim_fvu_drdm}.
A clear positive correlation can be observed in both panels, with the brightest stars having the smallest errors and FVU.

\section{Calibration Fiber}\label{sec:calib}

In transitioning from simulations to on-sky observations, we can first assess the ability of AIROPA to reconstruct the instrumental component of the PSF for the internal source described in Section~\ref{sec:inst}.
Here we present a summary of the analysis of the in-focus images of the calibration fiber, with the details reported in \citenum{bib:ciurlo21}.
Since the positioning of the mechanical arm holding the fiber and the light intensity are not very stable or repeatable, we do not know the intrinsic astrometry or photometry of the sources with a precision better than the measurements that we found in the previous sections.
We have instead relied on the study of the residuals , with the lower residuals indicating a more successful PSF-fitting.

The \texttt{variable-PSF} mode reduces the median FVU of the fiber images from $5.8\cdot10^{-3}$ to $2.4\cdot10^{-3}$, indicating a more accurate PSF than the \texttt{single-PSF}.
The number of speckles mistaken for real sources are also reduces by two thirds.

The instrumental phase maps measured in 2017 have also been tested with images of the fiber taken in 2018.
The use of the variable PSF yields the same improvement respect to the constant one as for the 2017 images, indicating a substantial stability with time of the instrumental aberrations.

\section{On Sky}\label{sec:sky}

\begin{figure}
\centering
\includegraphics[width=0.9\textwidth]{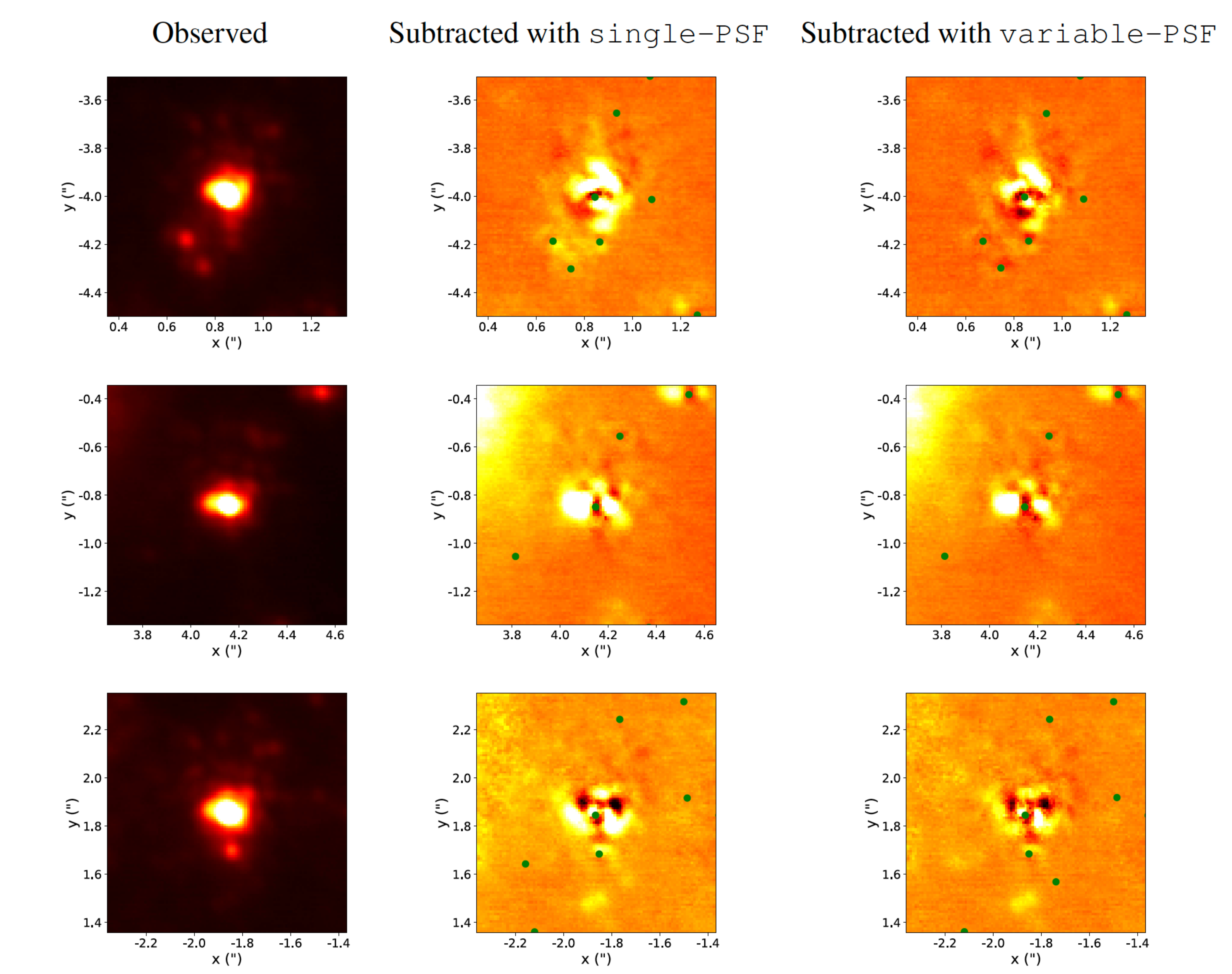}
\caption{Observed and residual images of several stars in the GC, from a NIRC2 exposure.
The \texttt{single-PSF} and \texttt{variable-PSF} subtracted images have the same color scale.
The green dots show the location where stars were detected.\label{fig:gc_obs_res_stars}}
\end{figure}

Analyzing the residual images of bright stars (Figure~\ref{fig:gc_obs_res_stars}), we notice that the \texttt{variable-PSF} is not performing distinctly better than with a homogeneous PSF.
A marginal reduction in the residual intensity can be observed, but not as clear as in simulations (Figure~\ref{fig:gc_sim_res_stars}).

\begin{figure}
\centering
\includegraphics[width=0.8\textwidth]{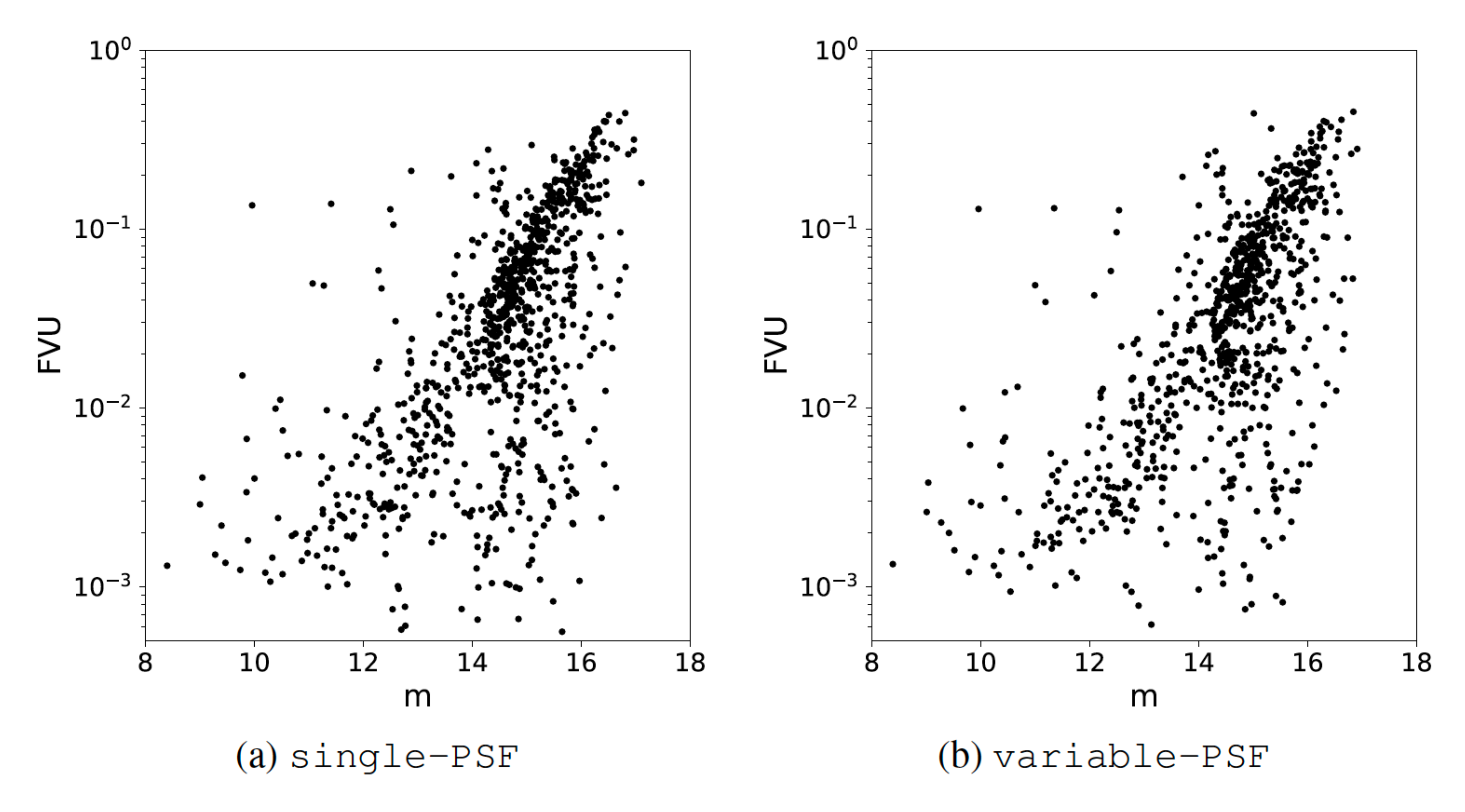}
\caption{FVU of the on-sky observations of the GC as a function of Kp instrumental magnitude.
The FVU of a star is its average on all exposures.\label{fig:gc_sky_fvu}}
\end{figure}

As with the calibration fiber images, we use the FVU to measure the the ability of AIROPA to reconstruct the PSF on GC images taken with NIRC2.
The average FVU was calculated for the stars that are identified in at least five of the 116 exposures of the GC field (Section~\ref{sec:nirc2img}).
The average FVU of stars brighter than $m_{Kp}=14$ mag does not improve by using the \texttt{variable-PSF} mode, changing to $3.6\cdot10^{-3}$ from $3.3\cdot10^{-3}$ of the \texttt{single-PSF} mode (Figure~\ref{fig:gc_sky_fvu}).

\begin{figure}
\centering
\includegraphics[width=0.6\textwidth]{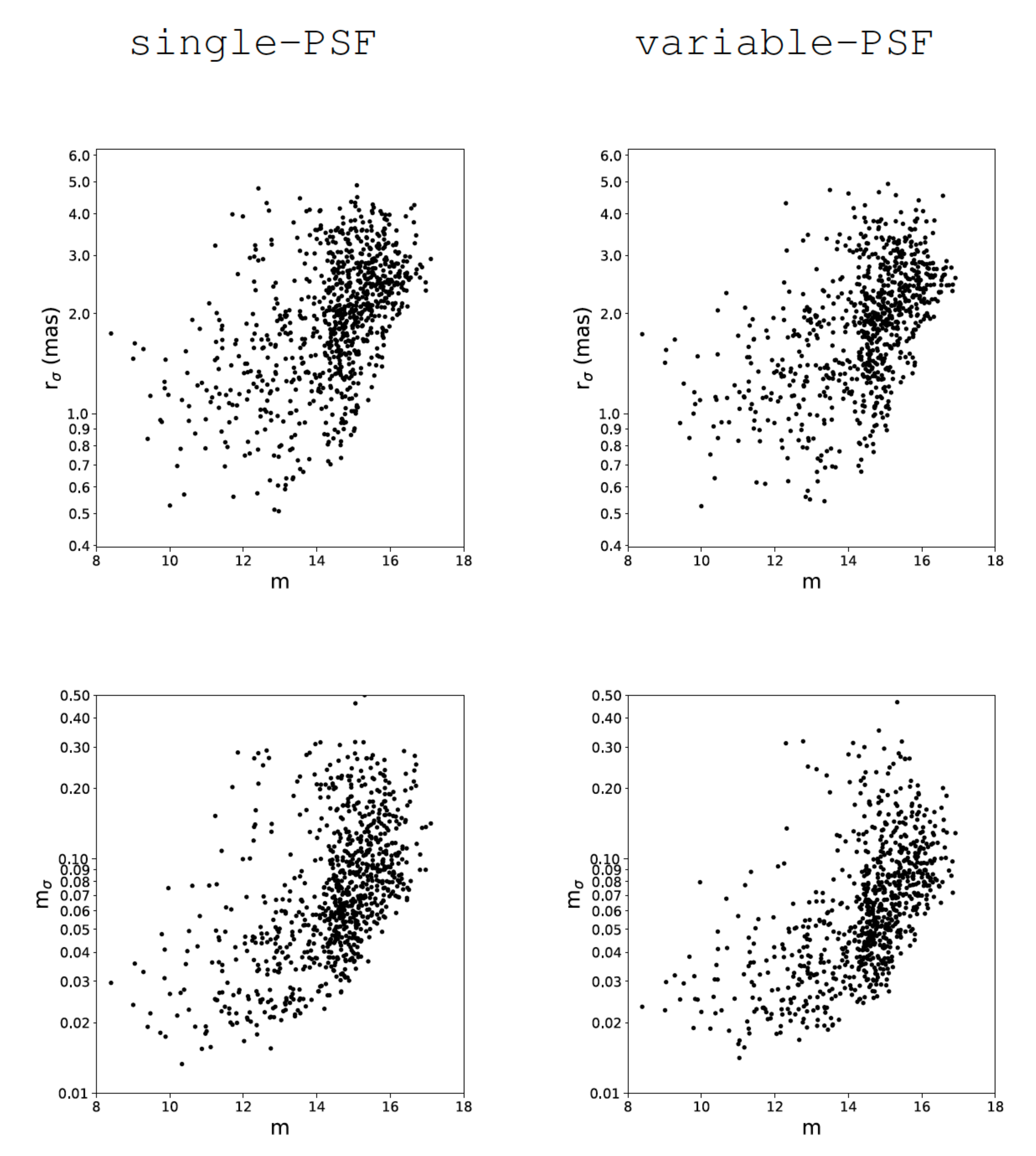}
\caption{Astrometric (first row) and photometric (second row) standard deviations with AIROPA on-sky GC images, as a function of Kp magnitude.\label{fig:gc_sky_res}}
\end{figure}

We can also measure the accuracy of the astrometry and photometry against the temporal variation of the PSF between images.
The smaller the standard deviation of the stars' positions and magnitudes, the better the ability of the reconstructed PSF to follow the change in PSF.
Average astrometric residuals of stars $m_{Kp}\leq 14$ have similar values between the \texttt{single-PSF} mode ($\sigma_{r}=1.29$ mas) to the \texttt{variable-PSF} mode ($\sigma_{r}=1.31$ mas) (top row of Figure~\ref{fig:gc_sky_res}).
A small improvement of 10\% is achieved with the average photometric residuals of bright stars, from $\sigma_{m_{Kp}}=3.4\cdot10^{-2}$ mag to $\sigma_{m_{Kp}}=3.1\cdot10^{-2}$ mag (bottom row of Figure~\ref{fig:gc_sky_res}).

\section{Discussion and Summary}\label{sec:summary}

We have tested the performance of AIROPA first with simulated NIRC2 images.
We have determined the baseline astrometric and photometric capabilities of the underlying fitting algorithm by analyzing a sparse fields of bright sources, to reduce the uncertainty caused by the shot noise, generated using a constant PSF.

To understand the effects of a PSF changing with direction, we have simulated another field of bright sources, but using a variable PSF with characteristics typical of NIRC2 observations of the GC.
Instrumental aberrations dominate over atmospheric angular anisoplanatism as cause in the spatial variation of the OTF and PSF (Figure~\ref{fig:otf}).
By using a sparse field, the systematic errors of PSF-fitting are caused only by the spatial variability and not by crowding.
We have demonstrated that the \texttt{variable-PSF} mode of AIROPA can reduce astrometric and photometric residuals on average by 80\%, respectively from $1.2$ to $1.9\cdot10^{-1}$ mas, and from $3.9\cdot10^{-2}$ mag to $6.1\cdot10^{-3}$ mag.

We have then replicated the effect of crowding on the measurements by simulating an image of the sources included in our catalog of the GC.
We find that for also for this case the \texttt{variable-PSF} mode provides strong gains in astrometry and photometry, with the average accuracy in position and magnitude changing from $8.3\cdot10^{-1}$ mas to $1.7\cdot10^{-1}$ mas and from $3.7\cdot10^{-2}$ mag to $0.6\cdot10^{-2}$ mag, respectively.
The improvement in the recovery of position and luminosity is more substantial for the most distant stars from the optical axis of the instrument, as can be seen by comparing the panels on the left and right of Figure~\ref{fig:var_rad}.

We have tested the capacity of AIROPA to reconstruct the instrumental part of the PSF by fitting the profile of the calibration fiber source with the predicted model (Section~\ref{sec:calib}).
Since we do not have a precise knowledge of the true position and luminosity of the source, we have used the FVU as diagnostic metric.
We have found that the \texttt{variable-PSF} of AIROPA provides a median FVU of $6.2\cdot10^{-5}$, still a significant advantage over the \texttt{single-PSF} mode (FVU=$5.7\cdot10^{-4}$).
Instrumental aberrations also present a remarkable stability, with the reconstructed PSF providing the same improvements in PSF fittings data taken after one year.

When we analyze real observations of the GC, we find that AIROPA does not improve the astrometry or photometry, with fitting residuals of bright stars similar between the \texttt{single-PSF} and \texttt{variable-PSF} modes.
The speckle patterns also change with the direction in the field of view.
To determine the cause of the difference between the predicted and observed PSF, we have considered the atmospheric and instrumental modules that constitute the PSF-reconstruction algorithm in AIROPA.

We believe that the source of the problem is not likely in the atmospheric simulation, since it does not dominate the reconstruction of the PSF shape and its variability, when compared to the contribution of the instrument (Figure~\ref{fig:otf_3}).
Also, our atmospheric modelling deals principally with the angular anisoplanatism effect on long exposure, and is not capable to reproduce the speckles like those observed in the residuals.

We also judge the phase diversity algorithm used for measuring the instrumental OTF to be sound, because of the positive results when tested with the calibration fiber images.
One important contribution to the total instrumental aberrations that is not accounted in our model is the impact of quasi-static aberrations introduced by the telescope.
Since they would be introduced before the calibration unit, they are not probed by the calibration fiber, and therefore are not included in our instrumental phase maps.
The 10 m primary mirror of the Keck telescopes requires a careful positioning of the segments \cite{bib:ragland18b}.
It is, however, unlikely that an error in the alignment caused by the telescope elevation is the source of the speckles observed, since the primary mirror is in a pupil, and they would appear with the same pattern across the field of view.
It is instead possible that the sagging of the secondary mirror or aberrations introduced by the telescope's K-mirror could be the cause of field-dependent aberrations that are not modeled and that dominate over the internal aberrations of the AO bench and NIRC2 that AIROPA is trying to suppress.

To overcome the shortfall of AIROPA with on-sky observation, our next step will be to measure static and quasi-static aberrations directly on sky with phase diversity, as in \citenum{bib:ragland18b}.
We will use a bright star with a similar elevation to the GC during our observations.
By dithering the telescope, we would also position it at different detector coordinates, to sample the PSF in different directions of the field of view.
By comparing our model of the wavefront with that measured on NIRC2 with phase diversity, we expect to identify the source of the described inconsistencies.
This necessary step will ultimately allow us to understand the aberrations of the full optical system, to correct the approach of AIROPA and to reliably use PSF reconstruction on sky.

\subsection*{Acknowledgments}

The data presented herein were obtained at the W. M. Keck Observatory, which is operated as a scientific partnership among the California Institute of Technology, the University of California and the National Aeronautics and Space Administration.
The Observatory was made possible by the generous financial support of the W. M. Keck Foundation.
The authors wish to recognize and acknowledge the very significant cultural role and reverence that the summit of Maunakea has always had within the indigenous Hawaiian community.
We are most fortunate to have the opportunity to conduct observations from this mountain.

We thank the staff of the Keck Observatory, especially Randy Campbell, Jim Lyke, Carlos \'{A}lvarez, and Peter Wizinowich, for all their help in obtaining calibration data.
We acknowledge support from the W. M. Keck Foundation, the Heising-Simons Foundation, the Gordon and Betty Moore Foundation, and the National Science Foundation (AST-1412615, AST-1518273).

Authors would like to thank the anonymous referee for helpful suggestions that improved the structure and readability of the paper.

\bibliography{biblio}

\begin{thebibliography}{10}

\bibitem{bib:davies12}
R.~Davies and M.~Kasper, ``Adaptive optics for astronomy,'' {\em ARA\&A} {\bf
  50}, 305  (2012).

\bibitem{bib:balick74}
B.~Balick and R.~L. Brown, ``Intense sub-arcsecond structure in the {G}alactic
  {C}enter,'' {\em ApJ} {\bf 194}, 265  (1974).

\bibitem{bib:ghez98}
A.~M. Ghez, B.~L. Klein, M.~Morris, {\em et~al.}, ``High proper-motion stars in
  the vicinity of {S}agittarius {A}\textsuperscript{*}: evidence for a
  supermassive black hole at the center of our {G}alaxy,'' {\em ApJ} {\bf 509},
  678  (1998).

\bibitem{bib:ghez00}
A.~M. Ghez, M.~Morris, E.~E. Becklin, {\em et~al.}, ``The accelerations of
  stars orbiting the {M}ilky {W}ay's central black hole,'' {\em ApJ} {\bf 407},
  349  (2000).

\bibitem{bib:genzel10}
R.~Genzel, F.~Eisenhauer, and S.~Gillessen, ``The {G}alactic {C}enter massive
  black hole and nuclear star cluster,'' {\em RvMP} {\bf 82}, 3121  (2010).

\bibitem{bib:ghez05}
A.~M. Ghez, S.~D. Hornstein, J.~R. Lu, {\em et~al.}, ``The first laser guide
  star adaptive optics observations of the {G}alactic {C}enter: Sgr
  {A}\textsuperscript{*}'s infrared color and the extended red emission in its
  vicinity,'' {\em ApJ} {\bf 635}, 1087  (2005).

\bibitem{bib:ghez08}
A.~M. Ghez, S.~Salim, N.~N. Weinberg, {\em et~al.}, ``Measuring distance and
  properties of the {M}ilky {W}ay's central supermassive black hole with
  stellar orbits,'' {\em ApJ} {\bf 689}, 1044  (2008).

\bibitem{bib:boehle16}
A.~Boehle, A.~M. Ghez, R.~Schödel, {\em et~al.}, ``An improved distance and
  mass estimate for {S}gr {A}\textsuperscript{*} from a multistar orbit
  analysis,'' {\em ApJ} {\bf 830}, 17  (2016).

\bibitem{bib:lu09}
J.~R. Lu, A.~M. Ghez, S.~D. Hornstein, {\em et~al.}, ``A disk of young stars at
  the {G}alactic {C}enter as determined by individual stellar orbits,'' {\em
  ApJ} {\bf 690}, 1463  (2009).

\bibitem{bib:do13}
T.~Do, G.~D. Martinez, S.~Yelda, {\em et~al.}, ``Three-dimensional stellar
  kinematics at the {G}alactic {C}enter: measuring the {N}uclear {S}tar
  {C}luster spatial density profile, black hole mass, and distance,'' {\em
  ApJL} {\bf 779}, L6  (2013).

\bibitem{bib:hees17}
A.~Hees, T.~Do, A.~M. Ghez, {\em et~al.}, ``Testing general relativity with
  stellar orbits around the supermassive black hole in our {G}alactic
  {C}enter,'' {\em PhRvL} {\bf 118}, 211101  (2017).

\bibitem{bib:schodel10}
R.~Schödel, ``Accurate photometry with adaptive optics in the presence of
  anisoplanatic effects with a sparsely sampled {PSF}. the {G}alactic {C}enter
  as an example of a challenging target for accurate {AO} photometry,'' {\em
  A\&A} {\bf 509}, A58  (2010).

\bibitem{bib:trippe10}
S.~Trippe, R.~Davies, F.~Eisenhauer, {\em et~al.}, ``High-precision astrometry
  with {MICADO} at the {E}uropean {E}xtremely {L}arge {T}elescope,'' {\em
  MNRAS} {\bf 402}, 1126  (2010).

\bibitem{bib:ascenso15}
J.~Ascenso, B.~Neichel, M.~Silva, {\em et~al.}, ``{PSF} reconstruction for {AO}
  photometry and astrometry,'' in {\em Proceedings of the Fourth AO4ELT
  Conference},   (2015).

\bibitem{bib:fritz15}
T.~K. Fritz, N.~Kallivayalil, E.~R. {Carrasco Damele}, {\em et~al.},
  ``Astrometry with {MCAO} at {G}emini and at {ELT}s,'' in {\em Proceedings of
  the Fourth AO4ELT Conference},   (2015).

\bibitem{bib:turri17}
P.~Turri, A.~W. McConnachie, P.~B. Stetson, {\em et~al.}, ``Optimal stellar
  photometry for multi-conjugate adaptive optics systems using science-based
  metrics,'' {\em AJ} {\bf 153}, 199  (2017).

\bibitem{bib:fusco00}
T.~Fusco, J.-M. Conan, L.~M. Mugnier, {\em et~al.}, ``Characterization of
  adaptive optics point spread function for anisoplanatic imaging.
  {A}pplication to stellar field deconvolution,'' {\em A\&AS} {\bf 142}, 149
  (2000).

\bibitem{bib:lamb14}
M.~Lamb, D.~R. Andersen, J.-P. V{\'e}ran, {\em et~al.}, ``Non-common path
  aberration corrections for current and future {AO} systems,'' in {\em
  Adaptive Optics Systems IV},  E.~Marchetti, L.~M. Close, and J.-P. V{\'e}ran,
  Eds., {\em Proceedings of the SPIE} {\bf 9148}, 914857  (2014).

\bibitem{bib:fried82}
D.~L. Fried, ``Anisoplanatism in adaptive optics,'' {\em JOSA} {\bf 72}, 52
  (1982).

\bibitem{bib:stetson87}
P.~B. Stetson, ``{DAOPHOT}: a computer program for crowded-field stellar
  photometry,'' {\em PASP} {\bf 99}, 191  (1987).

\bibitem{bib:schreiber12}
L.~Schreiber, E.~Diolaiti, A.~Sollima, {\em et~al.}, ``Developing a new
  software package for psf estimation and fitting of adaptive optics images,''
  in {\em Adaptive Optics Systems III},  B.~L. Ellerbroek, E.~Marchetti, and
  J.-P. V{\'e}ran, Eds., {\em Proceedings of the SPIE} {\bf 8447}, 84475V
  (2012).

\bibitem{bib:veran97}
J.-P. V{\'e}ran, F.~J. Rigaut, H.~Maitre, {\em et~al.}, ``Estimation of the
  adaptive optics long-exposure point-spread function using control loop
  data,'' {\em JOSAA} {\bf 14}, 3057  (1997).

\bibitem{bib:wagner19}
R.~Wagner, O.~Beltramo-Martin, C.~M. Correia, {\em et~al.}, ``Overview of {PSF}
  determination techniques for adaptive-optics assisted {ELT} instruments,'' in
  {\em Proceedings of the Sixth AO4ELT Conference},   (2019).

\bibitem{bib:martin16}
O.~A. Martin, C.~M. Correia, E.~Gendron, {\em et~al.}, ``{PSF} reconstruction
  validated using on-sky {CANARY} data in {MOAO} mode,'' in {\em Adaptive
  Optics Systems V},  E.~Marchetti, L.~M. Close, and J.-P. V{\'e}ran, Eds.,
  {\em Proceedings of the SPIE} {\bf 9909}, 99091Q  (2016).

\bibitem{bib:ragland18a}
S.~Ragland, T.~J. Dupuy, L.~Jolissaint, {\em et~al.}, ``Status of point spread
  function determination for {K}eck adaptive optics,'' in {\em Adaptive Optics
  Systems VI},  L.~M. Close, L.~Schreiber, and D.~Schmidt, Eds., {\em
  Proceedings of the SPIE} {\bf 10703}, 107031J  (2018).

\bibitem{bib:gilles18}
L.~Gilles, L.~Wang, and C.~Boyer, ``Point spread function reconstruction
  simulations for laser guide star multi-conjugate adaptive optics on extremely
  large telescopes,'' in {\em Adaptive Optics Systems VI},  L.~M. Close,
  L.~Schreiber, and D.~Schmidt, Eds., {\em Proceedings of the SPIE} {\bf
  10703}, 1070349  (2018).

\bibitem{bib:massari20}
D.~Massari, A.~Marasco, O.~Beltramo-Martin, {\em et~al.}, ``Successful
  application of {PSF-R} techniques to the case of the globular cluster {NGC}
  6121 ({M} 4),'' {\em A\&A} {\bf 634}, L5  (2020).

\bibitem{bib:witzel16}
G.~Witzel, J.~R. Lu, A.~M. Ghez, {\em et~al.}, ``The {AIROPA} software package:
  milestones for testing general relativity in the strong gravity regime with
  {AO},'' in {\em Adaptive Optics Systems V},  E.~Marchetti, L.~M. Close, and
  J.-P. V{\'e}ran, Eds., {\em Proceedings of the SPIE} {\bf 9909}, 990910
  (2016).

\bibitem{bib:ciurlo21}
A.~Ciurlo, P.~Turri, G.~Witzel, {\em et~al.}, ``{AIROPA II}: modeling
  instrumental aberrations for off-axis point spread functions in adaptive
  optics,''  (in prep.).

\bibitem{bib:terry21}
S.~K. Terry, J.~R. Lu, P.~Turri, {\em et~al.}, ``{AIROPA IV}: validation with
  various science cases,''  (in prep.).

\bibitem{bib:diolaiti00}
E.~Diolaiti, O.~Bendinelli, D.~Bonaccini, {\em et~al.}, ``Analysis of
  isoplanatic high resolution stellar fields by the {S}tar{F}inder code,'' {\em
  A\&AS} {\bf 147}, 335  (2000).

\bibitem{bib:britton06}
M.~C. Britton, ``The anisoplanatic point-spread function in adaptive optics,''
  {\em PASP} {\bf 118}, 885  (2006).

\bibitem{bib:wizinowich06}
P.~L. Wizinowich, D.~{Le Mignant}, A.~H. Bouchez, {\em et~al.}, ``The {W}. {M}.
  {K}eck {O}bservatory {L}aser {G}uide {S}tar {A}daptive {O}ptics {S}ystem:
  overview,'' {\em PASP} {\bf 118}, 297  (2006).

\bibitem{bib:yelda10}
S.~Yelda, J.~R. Lu, A.~M. Ghez, {\em et~al.}, ``Improving {G}alactic {C}enter
  astrometry by reducing the effects of geometric distortion,'' {\em ApJ} {\bf
  725}, 331  (2010).

\bibitem{bib:fowler90}
A.~M. Fowler and I.~Gatley, ``Demonstration of an algorithm for read-noise
  reduction in infrared arrays,'' {\em ApJ} {\bf 353}, L33  (1990).

\bibitem{bib:ciurlo18}
A.~Ciurlo, T.~Do, G.~Witzel, {\em et~al.}, ``Off-axis {PSF} reconstruction for
  integral field spectrograph: instrumental aberrations and application to
  {K}eck/{OSIRIS} data,'' in {\em Adaptive Optics Systems VI},  L.~M. Close,
  L.~Schreiber, and D.~Schmidt, Eds., {\em Proceedings of the SPIE} {\bf
  10703}, 107031O  (2018).

\bibitem{bib:sitarski14}
B.~N. Sitarski, G.~Witzel, M.~P. Fitzgerald, {\em et~al.}, ``Modeling
  instrumental field-dependent aberrations in the {NIRC}2 instrument on the
  {K}eck {II} telescope,'' in {\em Adaptive Optics Systems IV},  E.~Marchetti,
  L.~M. Close, and J.-P. V{\'e}ran, Eds., {\em Proceedings of the SPIE} {\bf
  9148}, 91486T  (2014).

\bibitem{bib:gerchberg72}
R.~W. Gerchberg and W.~O. Saxton, ``A practical algorithm for the determination
  of phase from image and diffraction plane pictures,'' {\em Optik} {\bf 35},
  237  (1972).

\bibitem{bib:gautam19}
A.~K. Gautam, T.~Do, A.~M. Ghez, {\em et~al.}, ``An adaptive optics survey of
  stellar variability at the {G}alactic {C}enter,'' {\em ApJ} {\bf 871}, 103
  (2019).

\bibitem{bib:jia19}
S.~Jia, J.~R. Lu, S.~Sakai, {\em et~al.}, ``The {G}alactic {C}enter: improved
  relative astrometry for velocities, accelerations, and orbits near the
  supermassive black hole,'' {\em ApJ} {\bf 873}, 9  (2019).

\bibitem{bib:service16}
M.~Service, J.~R. Lu, R.~Campbell, {\em et~al.}, ``A new distortion solution
  for {NIRC}2 on the {K}eck {II} telescope,'' {\em PASP} {\bf 128}, 095004
  (2016).

\bibitem{bib:rocca74}
A.~Rocca, F.~Roddier, and J.~Vernin, ``Detection of atmospheric turbulent
  layers by spatiotemporal and spatioangular correlation measurements of
  stellar-light scintillation,'' {\em JOSA} {\bf 64}, 1000  (1974).

\bibitem{bib:martin87}
H.~M. Martin, ``Image motion as a measure of seeing quality,'' {\em PASP} {\bf
  99}, 1360  (1987).

\bibitem{bib:sarazin90}
M.~Sarazin and F.~Roddier, ``The {ESO} differential image motion monitor,''
  {\em A\&A} {\bf 227}, 294  (1990).

\bibitem{bib:kornilov03}
V.~Kornilov, A.~A. Tokovinin, O.~Vozyakova, {\em et~al.}, ``{MASS}: a monitor
  of the vertical turbulence distribution,'' in {\em Adaptive Optical System
  Technologies II},  P.~L. Wizinowich and D.~Bonaccini, Eds., {\em Proceedings
  of the SPIE} {\bf 4839}, 837  (2003).

\bibitem{bib:tokovinin03}
A.~Tokovinin, V.~Kornilov, N.~Shatsky, {\em et~al.}, ``Restoration of
  turbulence profile from scintillation indices,'' {\em MNRAS} {\bf 343}, 891
  (2003).

\bibitem{bib:do18}
T.~Do, A.~Ciurlo, G.~Witzel, {\em et~al.}, ``Point-spread function
  reconstruction for integral-field spectrograph data,'' in {\em Adaptive
  Optics Systems VI},  L.~M. Close, L.~Schreiber, and D.~Schmidt, Eds., {\em
  Proceedings of the SPIE} {\bf 10703}, 107031I  (2018).

\bibitem{bib:ragland18b}
S.~Ragland, ``A novel technique to measure residual systematic segment piston
  errors of large aperture optical telescopes,'' in {\em Ground-based and
  Airborne Telescopes VII},  H.~K. Marshall and J.~Spyromilio, Eds., {\em
  Proceedings of the SPIE} {\bf 10700}, 107001D  (2018).

\end{thebibliography}
\bibliographystyle{spiejour}

\vspace{2ex}\noindent \textbf{Paolo Turri} is a Postdoctoral Researcher at University of British Columbia in the Department of Physics and Astronomy. Turri's interests are in adaptive optics, PSF reconstruction and AO science data analysis. Turri is currently involved in the design of the multi-object AO system for the GIRMOS instrument on Gemini North.

\vspace{1ex}\noindent \textbf{Jessica R. Lu} is an Associate Professor at University of California, Berkeley in the Astronomy Department. Lu’s expertise is in adaptive optics, astrometry, stellar populations and dynamics, gravitational microlensing, and black holes. Lu is the Project Scientist for adaptive optics projects including ‘imaka on the University of Hawaii 2.2 m telescope and KAPA at the W. M. Keck Observatory.

\vspace{1ex}\noindent Biographies of the other authors are not available.

\listoffigures
\end{document}